\documentclass{aa}
\usepackage[varg]{txfonts}
\usepackage{graphicx}
\usepackage{natbib}
\usepackage{color}
\bibpunct{(}{)}{;}{a}{}{,} 
\begin{document}
\title{Mesogranulation and small-scale dynamo action in the quiet Sun}
\author{P.~J.~Bushby\inst{\ref{inst1}} \and B.~Favier\inst{\ref{inst2}}}
\institute{School of Mathematics and Statistics, Newcastle University, Newcastle Upon Tyne, NE1 7RU, UK\\ \email{paul.bushby@ncl.ac.uk}\label{inst1}
\and
DAMTP, Centre for Mathematical Sciences, Cambridge University, Cambridge, CB3 0WA, UK\\ \email{bff21@cam.ac.uk}\label{inst2}}
\date{Received}
\abstract{Regions of quiet Sun generally exhibit a complex distribution of small-scale magnetic field structures, which interact with the near-surface turbulent convective motions. Furthermore, it is probable that some of these magnetic fields are generated locally by a convective dynamo mechanism. In addition to the well-known granular and supergranular convective scales, various observations have indicated that there is an intermediate scale of convection, known as mesogranulation, with vertical magnetic flux concentrations accumulating preferentially at the boundaries of mesogranules.}{Our aim is to investigate the small-scale dynamo properties of a convective flow that exhibits both granulation and mesogranulation, comparing our findings with solar observations.}{Adopting an idealised model for a localised region of quiet Sun, we use numerical simulations of compressible magnetohydrodynamics, in a three-dimensional Cartesian domain, to investigate the parametric dependence of this system (focusing particularly upon the effects of varying the aspect ratio and the Reynolds number).}{In purely hydrodynamic convection, we find that mesogranulation is a robust feature of this system provided that the domain is wide enough to accommodate these large-scale motions. The mesogranular peak in the kinetic energy spectrum is more pronounced in the higher Reynolds number simulations. We investigate the dynamo properties of this system in both the kinematic and the nonlinear regimes and we find that the dynamo is always more efficient in larger domains, when mesogranulation is present. Furthermore, we use a filtering technique in Fourier space to demonstrate that it is indeed the larger scales of motion that are primarily responsible for driving the dynamo. In the nonlinear regime, the magnetic field distribution compares very favourably to observations, both in terms of the spatial distribution and the measured field strengths.}{}
\maketitle

\section{Introduction}\label{sec:intro}

High resolution observations of the solar surface continue to provide new insights into the structure and evolution of small-scale magnetic fields in regions of quiet Sun. In any highly conducting fluid, overturning eddies tend to expel magnetic flux \citep[][]{weiss66}, so it is unsurprising that the magnetic field distribution at the solar surface is strongly influenced by the local convective motions. On the scale of the convectively-driven granulation (a typical granule has a width of approximately $1$Mm), the near-surface flow pattern naturally leads to the expulsion of vertical magnetic flux from the granular interiors, which then becomes concentrated into localised structures within the intergranular lanes \citep[see, for example,][]{linrimmele99}. Larger scales of convective motion also play an important role in determining the spatial distribution of quiet Sun magnetic fields. The boundaries of supergranules, which have typical diameters of approximately $30$Mm \citep[see e.g. ][]{rieutordrincon10}, are associated with a network of strong vertical magnetic flux concentrations \citep[][]{simonleighton64}. Furthermore, long-lived magnetic features in the internetwork regions tend to be advected preferentially towards the supergranular boundaries \citep[][]{wijnetal08,orozcoetal12}. Between the granular and the supergranular scales, many studies have also suggested the presence of mesogranular convective cells \citep[see, e.g.,][]{novemberetal81,mulleretal92,shineetal00,dominguez03,yellesetal11}. Whether or not mesogranulation is really an intrinsic scale of convection in the near-surface layers of the Sun is still a matter of  some debate \citep[see, e.g.,][]{rieutordetal10,rieutordrincon10,yellesetal11,katsukawaorozco12}. Nevertheless, observations of magnetic fields do seem to indicate that magnetic flux concentrations accumulate preferentially at mesogranular boundaries \citep[][]{dominguez03,wijnetal05,litesetal08,ishikawatsuneta11,yellesetal11}. 

\par In the quiet Sun, a near-surface magnetic structure with a characteristic field strength of approximately $400$G would have a magnetic energy density that is comparable in magnitude to the mean kinetic energy density of the surrounding non-magnetic convection \citep[][]{gallowayetal77}. It is perhaps surprising, therefore, that the peak vertical magnetic field strengths in regions of quiet Sun can often exceed a kG \citep[see, for example,][]{stenflo73,sanchezlites00,dominguezetal06b,dominguezetal06a,orozcoetal07}. The existence of such strong fields is usually attributed to a phenomenon known as convective collapse. This magnetic field intensification process is driven by vertical convective motions along the field lines, which naturally lead to the formation of partially-evacuated magnetic flux concentrations (within which the magnetic pressure is high enough to be comparable to the surrounding gas pressure). This phenomenon has been studied extensively within the context of simplified models of thin flux tubes \citep[see, for example,][]{webbroberts78,spruit79} as well as in more realistic magnetoconvection simulations \citep[][]{grossmannetal98,bushbyetal08}. Furthermore, recent Hinode observations seem to confirm that this is indeed the mechanism that is responsible for the production of kG-strength magnetic fields in regions of quiet Sun \citep[][]{nagataetal08}. In addition to these locally intense vertical magnetic flux concentrations, regions of quiet Sun also contain significant quantities of predominantly horizontal magnetic flux \citep[][]{orozcoetal07,litesetal08}. Indeed, within the field of view that was investigated by \citet{litesetal08}, the spatially averaged horizontal flux was found to be approximately five times larger than the corresponding (unsigned) mean vertical flux. Where the magnetic field is strongly inclined to the vertical (usually near the edges of granules), measurements indicate a typical field strength of the order of $100$G \citep[][]{orozcoetal07}. It is interesting to note that this value is comparable in magnitude to the estimated intrinsic field strength of the unresolved turbulent magnetic fields in these regions \citep[][]{trujilloetal04}.

\par Several studies have shown that magnetoconvection models can produce a magnetic field distribution that is similar to that observed in the quiet Sun \citep[][]{bushbyhoughton05,khomenkoetal05,steinnordlund06}. However, it has been suggested that a near-surface dynamo, driven by the local convective motions, is responsible for producing a significant part of the observed small-scale magnetic activity, and recent observations seem to support this idea \citep[see, e.g.,][]{buehleretal13}. Although solar-like parameter regimes cannot yet be studied (due to computational constraints), numerical simulations of convectively-driven dynamo action in a Cartesian domain can be regarded as an idealised representation of the dynamo process in a small region of quiet Sun. \citet{meneguzzipouquet89} and \citet{cattaneo99} established that non-rotating Boussinesq convection in an electrically-conducting fluid can act as an efficient dynamo provided that the magnetic Reynolds number is large. The resulting magnetic field distribution is highly disordered, although small-scale vertical magnetic flux concentrations do indeed accumulate within the convective downflows, as observed in the quiet Sun. A number of recent studies have investigated dynamo action in fully compressible convection. One approach to this problem is to simulate the dynamo in a model of quiet Sun convection which includes as many of the relevant physical processes as possible (such as radiative transfer and a realistic equation of state), and models of this type have produced results which compare favourably to observations \citep[][]{abbett07,voglerschussler07,schusslervogler08,pietarilaetal09,danilovicetal10,schussler13}. Other studies have considered dynamo action in much simpler models of compressible convection in a polytropic layer (or, in some cases, systems of more than one polytropic layers), including no non-essential physical effects \citep[][]{kapylaetal08,brummelletal10,bushbyetal10, bushbyetal11,bushbyetal12}. Minimal models of this form are well suited to parametric surveys. Furthermore, the findings of \citet{molletal11} suggest that additional physical processes such as radiative transfer probably do not play a major role in the operation of the dynamo, so it is instructive to study the properties of these comparatively simple models.  

\par Most previous simulations of dynamo action in compressible convection were carried out in relatively small computational domains, so are not able to investigate the effects of convective motions on scales that are larger than the basic granulation. In large domains, mesogranular cells have been observed in hydrodynamic Boussinesq convection \citep[][]{cattaneoetal01}. Furthermore, when this flow was allowed to drive a dynamo, regions of vertical magnetic flux were found to accumulate preferentially at the corners of mesogranules (as observed in the quiet Sun). Simulations of fully compressible convection have produced rather mixed results in terms of mesogranulation. Some hydrodynamic calculations have produced mesogranulation with a clearly-defined horizontal spatial scale \citep[see, for example,][]{rinconetal05}, whilst other studies have produced mesogranular cells with no intrinsic scale \citep[][]{matlochetal10}. In a very recent study, \citet{schussler13} found a mesogranular pattern whose scale was determined (at least partially) by the growth rate of the associated dynamo. So there are at least some suggestions in the literature that mesogranulation might be strongly model-dependent. This is clearly an area in which further study is needed.  

\par In a recent paper, \citet{bushbyetal12} investigated the dynamo properties of compressible convection in a large aspect ratio domain. Like \citet{rinconetal05} they found clear evidence for mesogranulation in their hydrodynamic flow, with a well-defined peak in the kinetic energy spectrum at the mesogranular scale. Furthermore, the calculations of \citet{bushbyetal12} suggest that the presence of mesogranulation is beneficial for dynamo action, at least in the kinematic regime (in comparable regions of parameter space, the dynamo is less efficient in domains that are too small to accommodate mesogranules). However, only a single flow was investigated in that study, so it is unclear how dependent these conclusions are upon the system parameters. In this paper, we carry out a systematic survey (subject to computational constraints) of the Reynolds number dependence of this system, focusing particularly upon the consequences for mesogranulation, before going on to consider the dynamo properties of the resultant flows. The model setup is described in the next section. We then discuss the hydrodynamic properties of the convective flows in Section 3, before moving on to describe the results from a series of dynamo calculations. Finally, our conclusions are presented in Section 4, where we discuss the relevance of our results to quiet Sun magnetism.   

\section{Model setup and governing equations}\label{sec:model}

The model is an idealised representation of convection in the near-surface layers of a region of quiet Sun. We consider the evolution of a layer of electrically-conducting, compressible fluid, which is heated from below. The fluid properties are characterised by various parameters, including the thermal conductivity, $K$, the shear viscosity, $\mu$, the magnetic diffusivity, $\eta$, the magnetic permeability, $\mu_0$, and the specific heat capacities at constant density and pressure, $c_v$ and $c_p$ (respectively). All of these quantities are assumed to be constant throughout the fluid, which occupies a Cartesian domain of dimensions $0 \le x,y \le \lambda d$ and $0 \le z \le d$. The axes are defined in such a way that the $z$-axis points vertically downwards, parallel to the constant gravitational acceleration, $\vec{g}=g\hat{\vec{z}}$, which implies that $z=0$ corresponds to the upper surface of the domain. Periodic boundary conditions are enforced in each of the horizontal directions, whilst (in this idealised model) the upper and lower bounding surfaces are assumed to be impermeable and stress-free. Both the upper and lower boundaries are held at fixed, uniform temperatures ($T_0$ and $T_0 + \Delta T$ respectively, with $\Delta T >0$). We also assume that the horizontal components of the magnetic field vector vanish at $z=0$ and $z=d$, which implies that the field is constrained to be vertical at the upper and lower boundaries.  

\begin{table*}
\begin{center}
\caption{ \label{table1}Specifying the simulation parameters for the convective flows.} 
\begin{tabular}{ccccccccccccc} 
\hline \hline
Run & $\lambda$ & Grid & $\gamma$ & $m$ & $\theta$ & $\kappa$ & $\sigma$ & $Ra$ & $Re$ &  $U_{\rm rms}$ & $t_{\rm conv}$ & $\rho_{\rm mid}$ \\ \hline
A1 & $4$ & $256\times 256 \times 96$ & $5/3$ & $1.0$ & $3.0$ & $0.00548$ & $1.5$ & $2.0 \times 10^5$ & $93$ & $0.315$ & $3.17$ & $2.418$\\
A2 & $4$ & $256\times 256 \times 96$ & $5/3$ & $1.0$ & $3.0$ & $0.00548$ & $1.0$ & $3.0 \times 10^5$ & $160$ & $0.363$ & $2.75$ & $2.416$\\
A3 & $4$ & $256\times 256 \times 120$ & $5/3$ & $1.0$ & $3.0$ & $0.00548$ & $0.5$ & $6.0 \times 10^5$ & $344$ & $0.390$ & $2.56$ & $2.416$\\
A4 & $4$ & $256\times 256 \times 120$ & $5/3$ & $1.0$ & $3.0$ & $0.00548$ & $0.375$ & $8.0 \times 10^5$ & $483$ & $0.411$ & $2.43$ & $2.416$\\ 
\hline
B1 & $10$ & $512\times 512 \times 96$ & $5/3$ & $1.0$ & $3.0$ & $0.00548$ & $1.5$ & $2.0 \times 10^5$ & $95$ & $0.322$ & $3.10$ & $2.420$\\
B2 & $10$ & $512\times 512 \times 96$ & $5/3$ & $1.0$ & $3.0$ & $0.00548$ & $1.0$ & $3.0 \times 10^5$ & $156$ & $0.353$ & $2.83$ & $2.417$\\
B3 & $10$ & $512\times 512 \times 120$ & $5/3$ & $1.0$ & $3.0$ & $0.00548$ & $0.5$ & $6.0 \times 10^5$ & $350$ & $0.397$ & $2.52$ & $2.416$\\
B4 & $10$ & $512\times 512 \times 120$ & $5/3$ & $1.0$ & $3.0$ & $0.00548$ & $0.375$ & $8.0 \times 10^5$ & $485$ & $0.413$ & $2.42$ & $2.414$\\ 
\hline
C1 & $20$ & $512\times 512 \times 96$ & $5/3$ & $1.0$ & $3.0$ & $0.00548$ & $1.5$ & $2.0 \times 10^5$ & $95$ & $0.322$ & $3.10$ & $2.421$\\
C2 & $20$ & $512\times 512 \times 96$ & $5/3$ & $1.0$ & $3.0$ & $0.00548$ & $1.0$ & $3.0 \times 10^5$ & $156$ & $0.353$ & $2.83$ & $2.419$\\
\hline
\end{tabular}
\tablefoot{The parameters are as defined in the text -- the quoted values for $Re$, $U_{\rm rms}$, $t_{\rm conv}$, $\rho_{\rm mid}$ have all been time-averaged. Grid resolutions are given in the form $N_x \times N_y \times N_z$. Some of the higher $Rm$ dynamo calculations are carried out on a  $2N_x \times 2N_y \times N_z$ grid (where necessary, the interpolation of the flow onto a finer mesh is carried out in Fourier space before the magnetic field is introduced).}
\end{center}
\end{table*} 

\par Under these assumptions, the governing equations for the density $\rho(\vec{x},t)$, the fluid velocity $\vec{u}(\vec{x},t)$, the temperature $T(\vec{x},t)$ and magnetic field $\vec{B}(\vec{x},t)$ can be written in the following dimensionless form \citep[see, for example,][for more details]{bushbyetal08}:  

\begin{eqnarray}
\label{eq:mass}
&&\hspace{-0.3in}\frac{\partial \rho}{\partial t}=-\vec{u}\vec{\cdot}\vec{\nabla}\rho- \rho \vec{\nabla} \vec{\cdot} \vec{u},\\
\label{eq:momentum}
&&\hspace{-0.3in}\frac{\partial \vec{u}}{\partial t}=-\vec{u}\vec{\cdot}\vec{\nabla}\vec{u}-\frac{1}{\rho}\vec{\nabla}P+\frac{1}{\rho}\left(\nabla\times\vec{B}\right)\times\vec{B}+\theta(m+1)\hat{\vec{z}}+\frac{\kappa \sigma}{\rho}\vec{\nabla}\vec{\cdot}\tens{\tau},\\
\label{eq:heateq}
&&\hspace{-0.3in}\frac{\partial T}{\partial t}=-\vec{u}\vec{\cdot}\vec{\nabla} T - \left(\gamma -1\right)T\vec{\nabla} \vec{\cdot} \vec{u}+
\frac{\kappa\gamma}{\rho}\nabla^2 T \\ \nonumber &&\hspace{0.1in} +\frac{\kappa(\gamma-1)}{\rho}\left(\sigma |\tens{\tau}|^2/2 + \zeta_0|\vec{\nabla}
\times \vec{B}|^2\right),\\
\label{eq:induction}
&&\hspace{-0.3in}\frac{\partial \vec{B}}{\partial t}=\vec{\nabla} \times \left( \vec{u}\times \vec{B} -  \kappa \zeta_0 \vec{\nabla} \times \vec{B} \right),
\end{eqnarray} 

\noindent where the pressure $P(\vec{x},t)$ satisfies the equation of state for a perfect gas, $P=\rho T$, the components of the rate of strain tensor, $\tens{\tau}$, are defined by  

\begin{equation}
\label{eq:ros}
\tau_{ij} = \frac{\partial u_i}{\partial x_j}+\frac{\partial u_j}{\partial x_i}- \frac{2}{3}\delta_{ij} \frac{\partial u_k}{\partial x_k},
\end{equation} 

\noindent whilst the magnetic field satisfies $\nabla\cdot\vec{B}=0$. Note that all lengths have been scaled by $d$, which implies that the domain has unit depth in this dimensionless system, whilst $d/(R_*T_0)^{1/2}$ has been used to scale time. The temperature has been scaled by $T_0$, whilst the density of the fluid has been scaled by $\rho_0$ (the density at the upper surface in the absence of any motion). Finally, the velocity field has been scaled by $\left(R_* T_0\right)^{1/2}$, whilst the magnetic field has been scaled by $\left(\mu_0 R_* \rho_0T_0\right)^{1/2}$. In addition to the aspect ratio, $\lambda$, the evolution of the system is then determined by various dimensionless parameters. The ratio of specific heat capacities is defined by $\gamma=c_p/c_v$, whilst $m=\left(gd/R_*\Delta T\right) -1$ is the polytropic index. The temperature difference across the domain is determined by the stratification parameter, $\theta=\Delta T/T_0$. Defining $\kappa=K/d\rho_0 c_p \left(R_* T_0\right)^{1/2}$ to be the dimensionless thermal diffusivity, the remaining parameters are the Prandtl number, $\sigma=\mu c_p/K$, and $\zeta_0=\eta c_p \rho_0/K$, which represents the ratio of the magnetic to the thermal diffusivities at the upper surface of the domain.  

\par Given the complexity of the governing equations, we solve these numerically using an updated version of a well-tested code \citep[][]{matthewsetal95}. Horizontal derivatives are evaluated in Fourier space, whilst a fourth-order finite difference scheme is used to approximate the vertical derivatives. The equations are evolved in time using an explicit third-order Adams-Bashforth scheme with a variable timestep. A poloidal-toroidal decomposition is used to ensure that the magnetic field remains solenoidal. The code is parallelised using MPI, which enables us to investigate convectively-driven dynamos in large computational domains. In the absence of a magnetic field, the governing equations have an equilibrium solution, given by 

\begin{equation}
\label{eq:polytrope}
T = 1+\theta z, \; \rho = \left(1+\theta z\right)^m, \; \vec{u}=\vec{0},
\end{equation}
\noindent which corresponds to a hydrostatic polytropic layer. All of the hydrodynamic simulations in this paper are initialised by perturbing this equilibrium state via the introduction of a small, random perturbation to the temperature distribution.  

\section{Numerical results}\label{sec:results}

\begin{figure*}[t]
\begin{center}
\includegraphics[width=9cm]{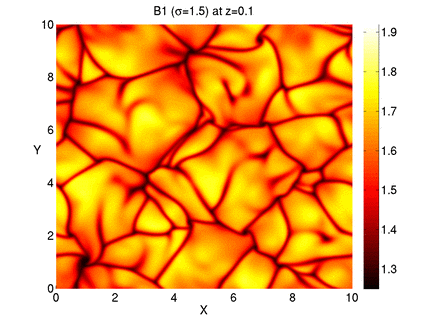}
\includegraphics[width=9cm]{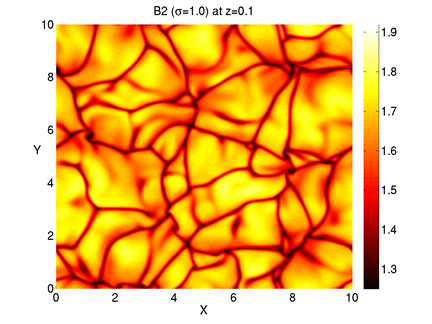}
\includegraphics[width=9cm]{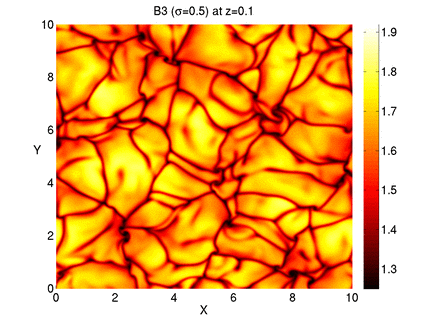}
\includegraphics[width=9cm]{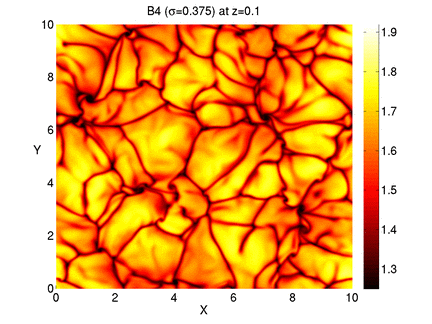}
\caption{Temperature distribution in the horizontal plane defined by $z=0.1$, for the $\lambda=10$ cases: B1 (top left), B2 (top right), B3 (bottom left) and B4 (bottom right).}
\label{fig1}
\end{center}
\end{figure*} 

\subsection{Specifying the model parameters}

Throughout this study, we fix the ratio of specific heat capacities to be $\gamma=5/3$, which is the appropriate value for a monatomic gas. Fixing the polytropic index to be $m=1$ implies that the layer is superadiabatically stratified, whilst a thermal stratification of $\theta=3$ implies that the temperature varies by a factor of $4$ across the domain. We choose a fixed value of $\kappa=0.00548$, whilst $\sigma$ varies in the range $0.375 \le \sigma \le 1.5$. These particular parameter choices imply that the mid-layer Rayleigh number 

\begin{equation}
\label{eq:rayl}
Ra = (m+1-m\gamma)\left(1+\theta/2\right)^{2m-1}\frac{(m+1)\theta^2}{\kappa^2\gamma\sigma},
\end{equation}
\noindent takes values in the range $2 \times 10^5 \le Ra \le 8 \times 10^5$. In a large aspect ratio domain, the critical Rayleigh number for the onset of convection in the absence of a magnetic field is $Ra_{\rm crit} \approx 900$, so these parameter choices ensure that the resulting convection will be vigorous and time-dependent. The Reynolds number of the system is a measurable parameter which provides some indication of the level of turbulence within the flow. Using the depth of the domain (which, as we have already stated, equals unity in these dimensionless units) as the characteristic length-scale, we define the mid-layer Reynolds number to be  

\begin{equation}
\label{eq:reynolds}
Re = \frac{\rho_{\rm mid}U_{\rm rms}}{\kappa \sigma},
\end{equation}

\noindent where $U_{\rm rms}$ is the total rms velocity, and $\rho_{\rm mid}$ is the mean density at the mid-layer of the domain. When considering the dynamo properties of this system, we shall often refer to the magnetic Reynolds number, which is defined by  

\begin{equation}
\label{eq:magreynolds}
Rm = \frac{U_{\rm rms}}{\kappa \zeta_0},
\end{equation}
\noindent whilst the (mid-layer) magnetic Prandtl number is defined to be  
 
\begin{equation}
\label{eq:magprandtl}
Pm = Rm/Re. 
\end{equation}

\noindent For a given flow, both $Rm$ and $Pm$ can be varied simultaneously by changing the value of $\zeta_0$. Dynamo action is expected only in the high $Rm$ (low $\zeta_0$) regime. Finally, following \citet{cattaneo99}, we define the convective turnover time to be $t_{\rm conv} = 1/U_{\rm rms}$.  

\par The hydrodynamic parameters for the simulation set that will be considered in this paper are defined in Table~\ref{table1}. In order to investigate the dependence of the flows upon the size of the computational domain, calculations have been carried out for $\lambda=4$, $\lambda=10$ and $\lambda=20$. Although the flows in cases C1 and C2 are adequately resolved for these values of $\sigma$, it was not possible to carry out $\lambda=20$ calculations at lower $\sigma$ (and consequently higher $Re$) due to computational constraints. For a given value of $\sigma$, it is clear that $Re$ and $t_{\rm conv}$ are only weakly dependent upon $\lambda$. Indeed, in this respect, the $\lambda=10$ and $\lambda=20$ cases are remarkably similar, whilst the small differences in the $\lambda=4$ case suggest that the corresponding flow is weakly-constrained by the size of the domain in the $\lambda=4$ calculations. Of course, this is a very crude comparison, so further analysis is needed in order to analyse the extent to which these flows are dependent upon the choice of aspect ratio.   

\subsection{Granulation and mesogranulation in hydrodynamic convection}\label{subsec:meso}

\begin{figure}[t]
\begin{center}
\includegraphics[width=9.0cm]{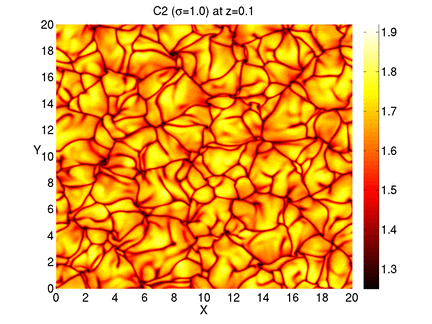}\\
\includegraphics[width=9.0cm]{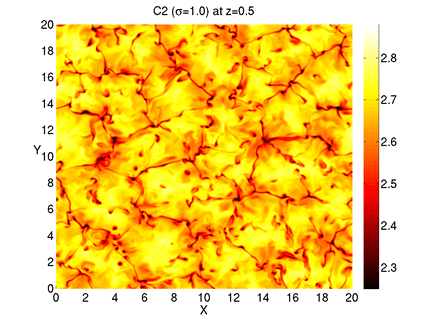}
\caption{Temperature distribution in two different horizontal planes for the $\lambda=20$, $\sigma=1.0$ case (C2). The upper plot shows the $z=0.1$ plane, whilst the lower plot shows the $z=0.5$ plane.} 
\label{fig2}
\end{center}
\end{figure}

The convective flows for the $\lambda=10$ cases (B1 -- B4) are illustrated in Figure~\ref{fig1}. All simulations produce highly time-dependent flows, with a near-surface pattern of convection that is comparable to the granular pattern that is observed at the surface of the Sun (with broad, warm upflows surrounded by a network of cooler, narrow downflows). In qualitative terms, it is clear that there tends to be more small-scale structure in the granular pattern at higher Reynolds numbers.  Figure~\ref{fig2} shows the temperature distribution in two different horizontal planes for one of the $\lambda=20$ cases (C2). In the $z=0.1$ plot in the upper part of Figure~\ref{fig2}, the near-surface granulation pattern is qualitatively similar to that observed in the corresponding $\lambda=10$ case (B2). However there is some evidence for the existence of structure on larger scales in the temperature distribution at the mid-plane. Only the strongest convective downflows from the surface layers penetrate to this depth -- if we were to argue that these strong downflows correspond to the boundaries of mesogranules, then this mid-plane temperature distribution would provide some visual indication of the corresponding mesogranular scale. However, such an assertion clearly needs to be tested by carrying out a more quantitative analysis of these hydrodynamic flows.

\par
\begin{figure}[t]
\begin{center}
\includegraphics[width=9.0cm]{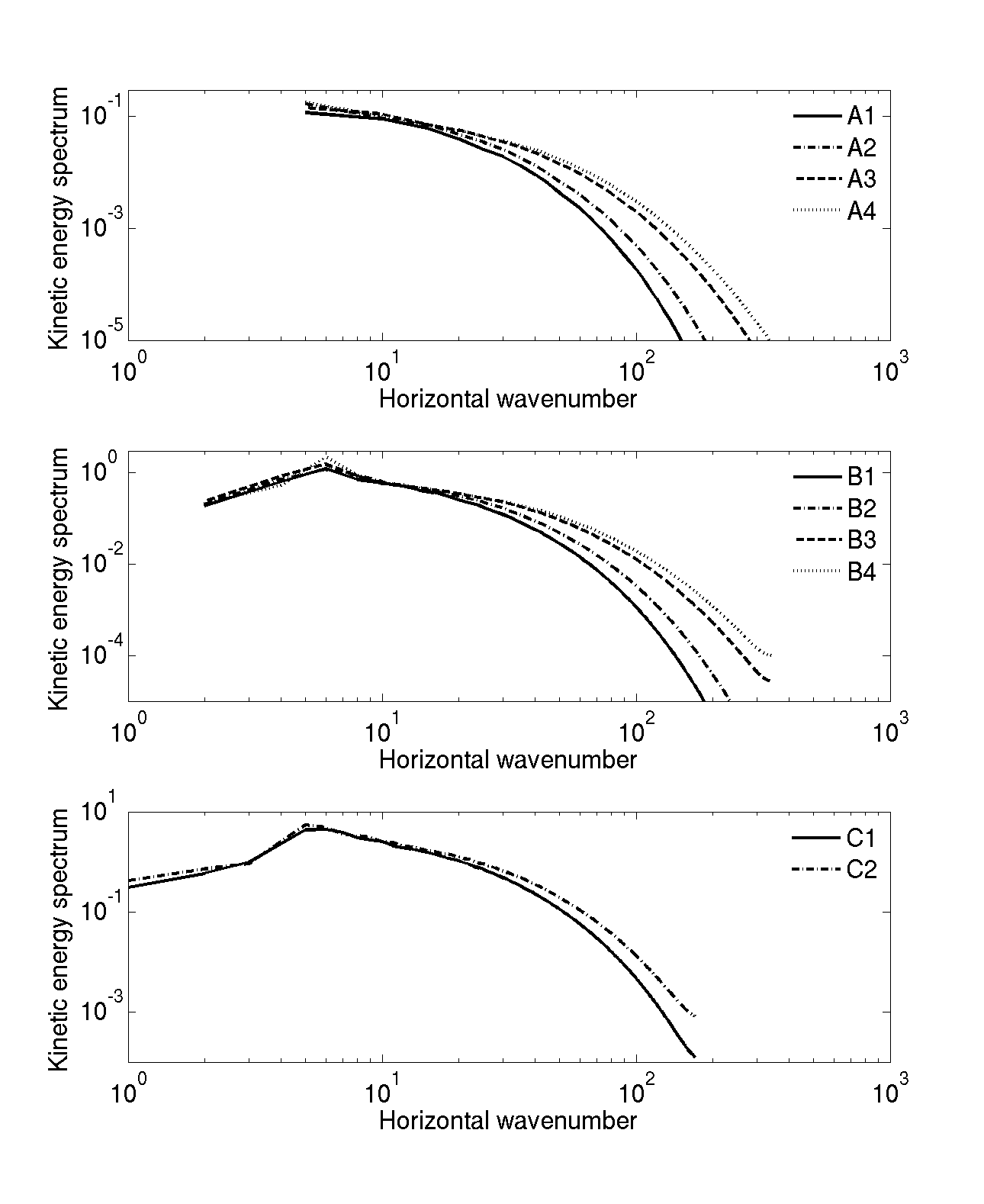}
\caption{Kinetic energy spectra as a function of the horizontal wavenumber for the hydrodynamic convective flows that are defined in Table~\ref{table1}. Top: $\lambda=4$; Middle: $\lambda=10$; Bottom: $\lambda=20$. In the upper two plots, the wavenumber has been normalised so that the scales are comparable in all cases.}
\label{fig3}
\end{center}
\end{figure}

 To determine conclusively whether or not mesogranulation is present in these simulations, we calculated the time- and depth-averaged kinetic energy spectra, based upon the horizontal wavenumber, $k_{H}$. These kinetic energy spectra are shown in Figure~\ref{fig3}, where the wavenumber in the lower aspect ratio cases has been scaled for consistency with the corresponding $\lambda=20$ spectra. In a previous paper \citep[][]{bushbyetal12} we investigated the depth-dependence of the kinetic energy spectra for cases A2 and B2, concluding that signatures of mesogranulation (if it is present) can be seen at all depths. We therefore decided to focus upon depth-averaged spectra here to simplify the analysis. Various conclusions can be drawn from these kinetic energy spectra. Firstly, there is a clear ``mesogranular'' peak at low wavenumber ($k_H \approx 5-6$) in all of the $\lambda=10$ and $\lambda=20$ cases. This corresponds to the horizontal scale that can be identified visually in the mid-plane temperature distribution that is illustrated in the lower part of Figure~\ref{fig2}. Furthermore, although it is present for all Reynolds numbers, the peak at this mesogranular scale seems to be more pronounced in the higher $Re$ cases, which (unsurprisingly) also exhibit more structure at higher $k_H$. Regardless of the value of $Re$, it is clear that the flow is constrained by the size of the box in the $\lambda=4$ cases, where the width of the domain is comparable to the mesogranular scale. There are no obvious indications of geometrical constraints in any of the other cases. Indeed, given the similarities between the comparable spectra in the $\lambda=10$ and $\lambda=20$ cases, we would argue that a $\lambda=10$ domain should be large enough for the purposes of this particular study. We shall therefore not consider the dynamo properties of the $\lambda=20$ simulations in this paper, deferring all further considerations of this case to a future study.  

\par
\begin{figure}[t]
\begin{center}
\includegraphics[width=9.0cm]{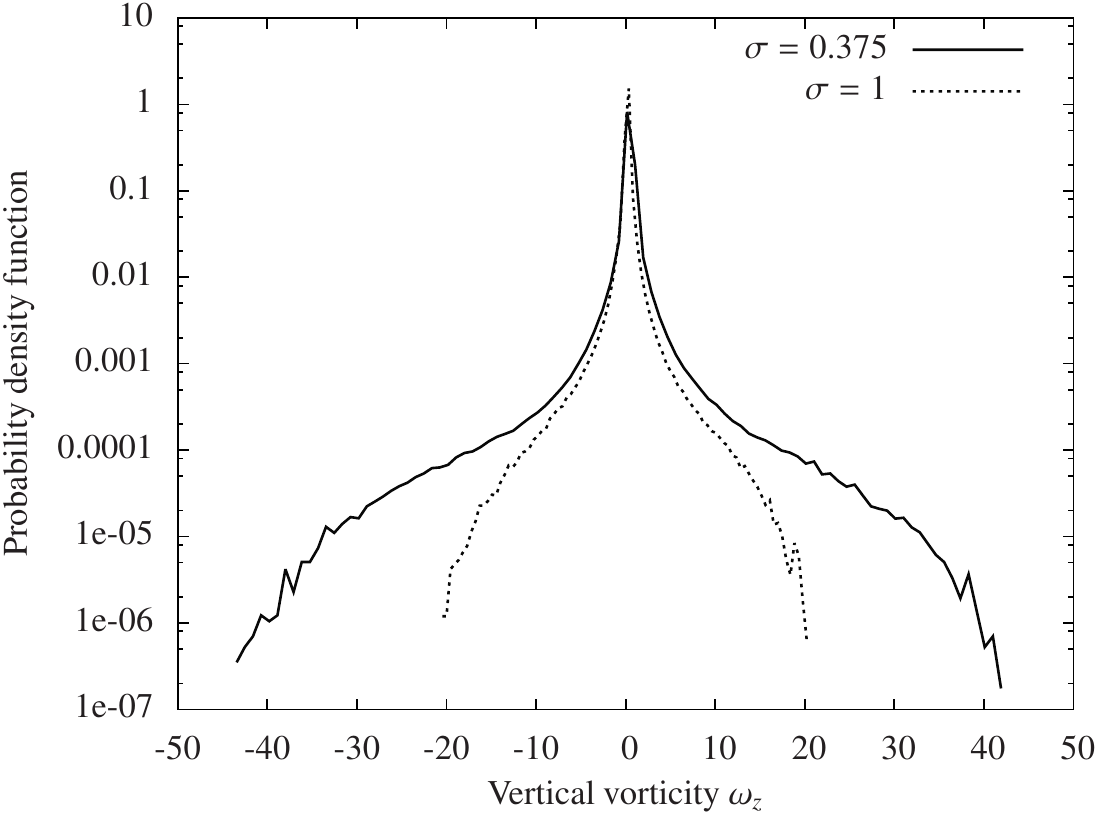}
\caption{Time-averaged probability density function of the vertical vorticity close to the upper boundary ($z=0.1$) for simulations B2 (dotted line) and B4 (solid line).}
\label{fig4}
\end{center}
\end{figure}

\begin{figure}[t]
\begin{center}
\includegraphics[width=9.0cm]{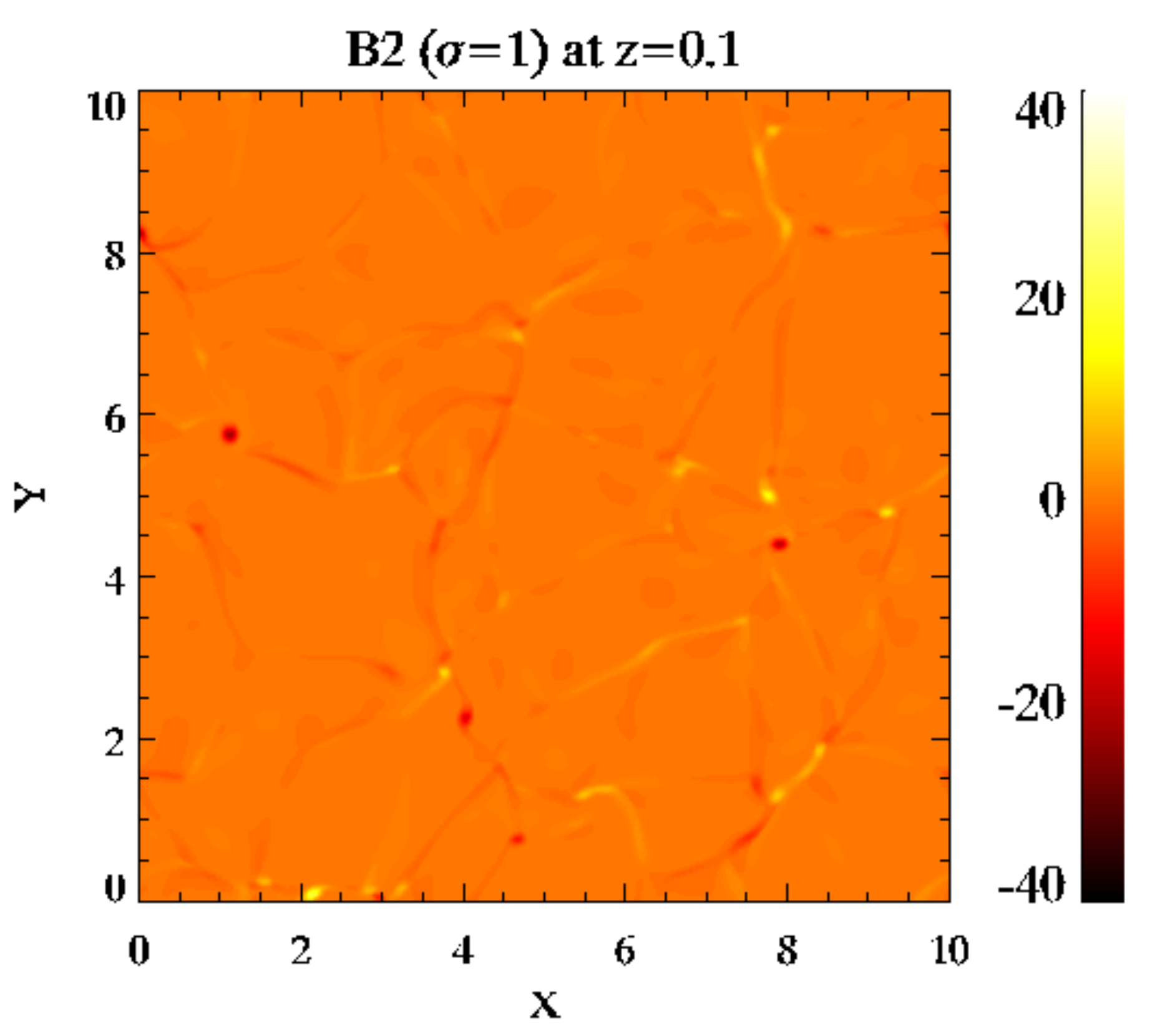}\\
\includegraphics[width=9.0cm]{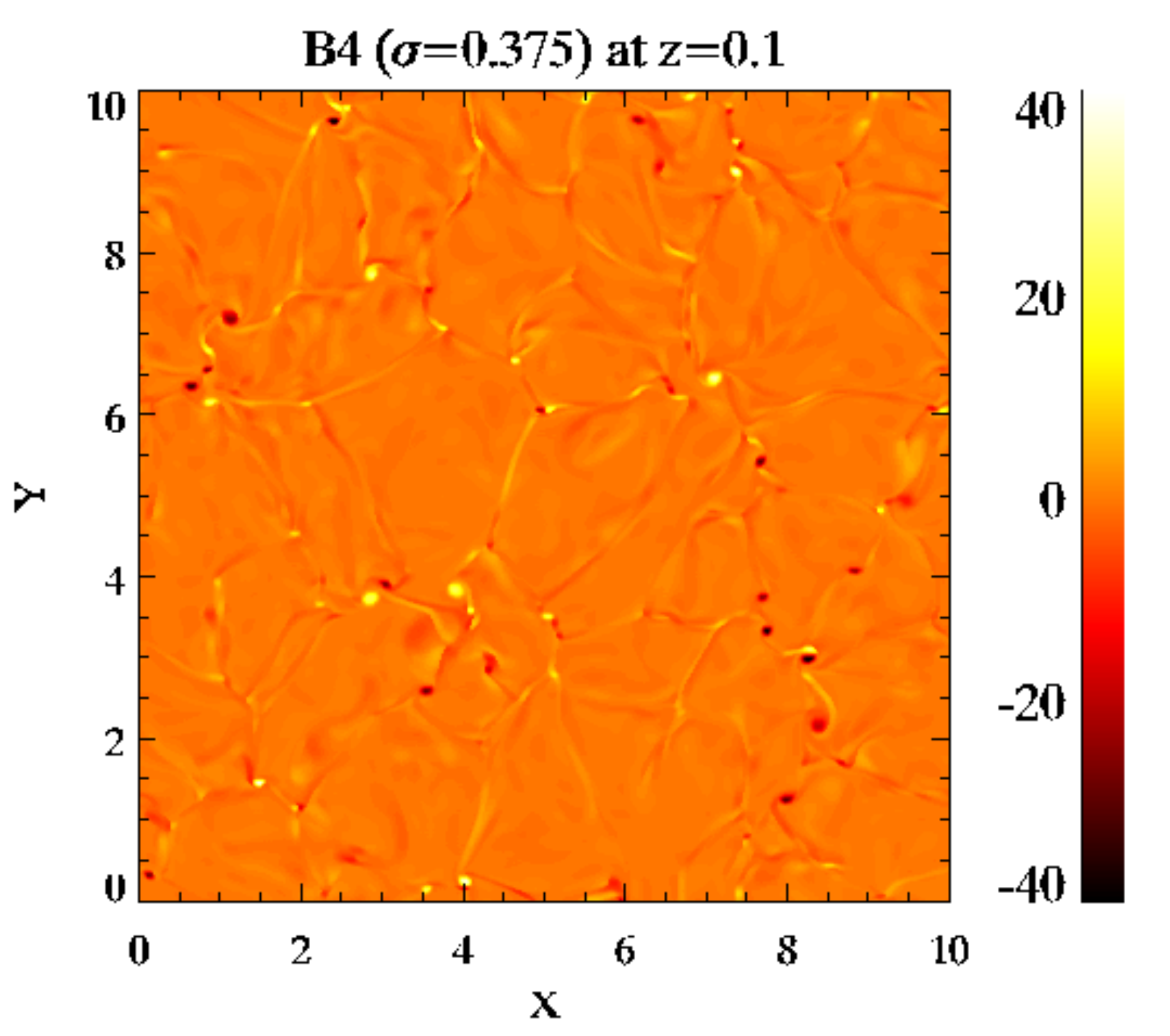}
\caption{Vertical vorticity distribution in the $z=0.1$ horizontal plane. The upper plot corresponds to simulation B2, whilst the lower plot shows the results from simulation B4.} 
\label{fig5}
\end{center}
\end{figure}

 Mesogranulation is clearly a robust feature of this system, which raises questions about its origin and stability. Even though the peak in the kinetic energy energy spectrum is more pronounced in the higher Reynolds number cases, it is also intriguing to note that the horizontal scale of the mesogranular cells seems to be only rather weakly-dependent upon $Re$. Whilst we are not yet able to speculate about how this horizontal scale is determined, differences in the behaviour of the vertical vorticity across the various simulations may provide some clues regarding the origin of mesogranulation \citep[and it has already been suggested that the vorticity distribution may be playing an important role in this regard, see e.g.][]{cattaneoetal01}. We show in Figure~\ref{fig4} the time-averaged probability density function of the vertical component of the vorticity in the horizontal plane at $z=0.1$, for simulations B2 and B4, whilst Figure~\ref{fig5} shows a snapshot of the distribution of the vertical vorticity in the same horizontal plane for the same two cases. In the higher $Re$ case (B4) rare events with very large absolute values of the vertical vorticity can be seen in the probability density function. The strongest localised vortices that can be seen in the lower part of Figure~\ref{fig5} are coincident with the corners of the mesogranules \citep[something that can easily be verified by using a passive tracer method, see e.g.][]{bushbyetal12}, so we would argue that it is probable that these features play a role in the formation and maintenance of the mesogranular pattern. If this causal connection does indeed exist, the fact that these concentrated vortices are located at the edges of the mesogranules, where large-scale straining motions can amplify them, could explain why mesogranules persist for many turnover times. The weaker vortices that can be seen in the B2 case would also then be consistent with the fact that the mesogranulation is less pronounced in this lower $Re$ case. It is worth noting here that the horizontal components of the vorticity seem to exhibit a much weaker dependence upon the Reynolds number, so the observed $Re$-dependence in the mesogranular pattern is more likely to be attributable to changes in the vertical vorticity profile. The dependence of the vorticity dynamics upon the aspect ratio also tends to support this hypothesis: Presumably as a consequence of geometric constraints, wide tails in the probability density function for the vertical vorticity are only seen in the large $\lambda$ cases in which mesogranulation is observed.

\subsection{Dynamo action in the kinematic regime}\label{subsec:kindynamo}

Having investigated the hydrodynamic properties of these flows, we now turn our attention to their dynamo properties. We start off each dynamo calculation by introducing a seed magnetic field into fully-developed hydrodynamic convection. This seed field is initially of low amplitude and is purely vertical, with a high wavenumber sinusoidal dependence upon both $x$ and $y$. Whether the total magnetic energy grows or decays is then determined uniquely by the value of $\zeta_0$ (or, equivalently, $Rm$), which can be specified as the field is introduced. For low $Rm$ (high $\zeta_0$), the evolution of the field is dominated by effects of ohmic dissipation, which implies that there is no dynamo. Dynamo action only becomes possible when the magnetic Reynolds number exceeds some critical value, $Rm_{\rm crit}$. In the dynamo regime, the magnetic energy can exhibit strong fluctuations although, whilst the field is still weak, the overall trend is for exponential growth. In numerical simulations, this phase of kinematic growth can be prolonged indefinitely by artificially removing the Lorentz force terms from the system of equations. Here, we adopt this approach to determine how $Rm_{\rm crit}$ (and how the $Rm$-dependence of the dynamo growth rate) depends upon the other parameters in the system. Particularly close to  $Rm_{\rm crit}$, where there can be significant long-term fluctuations in the magnetic energy, it is necessary to evolve these calculations over many convective turnover times in order to obtain reliable estimates for the kinematic growth rates. 

\begin{figure}[h]
\begin{center}
\includegraphics[width=9.0cm]{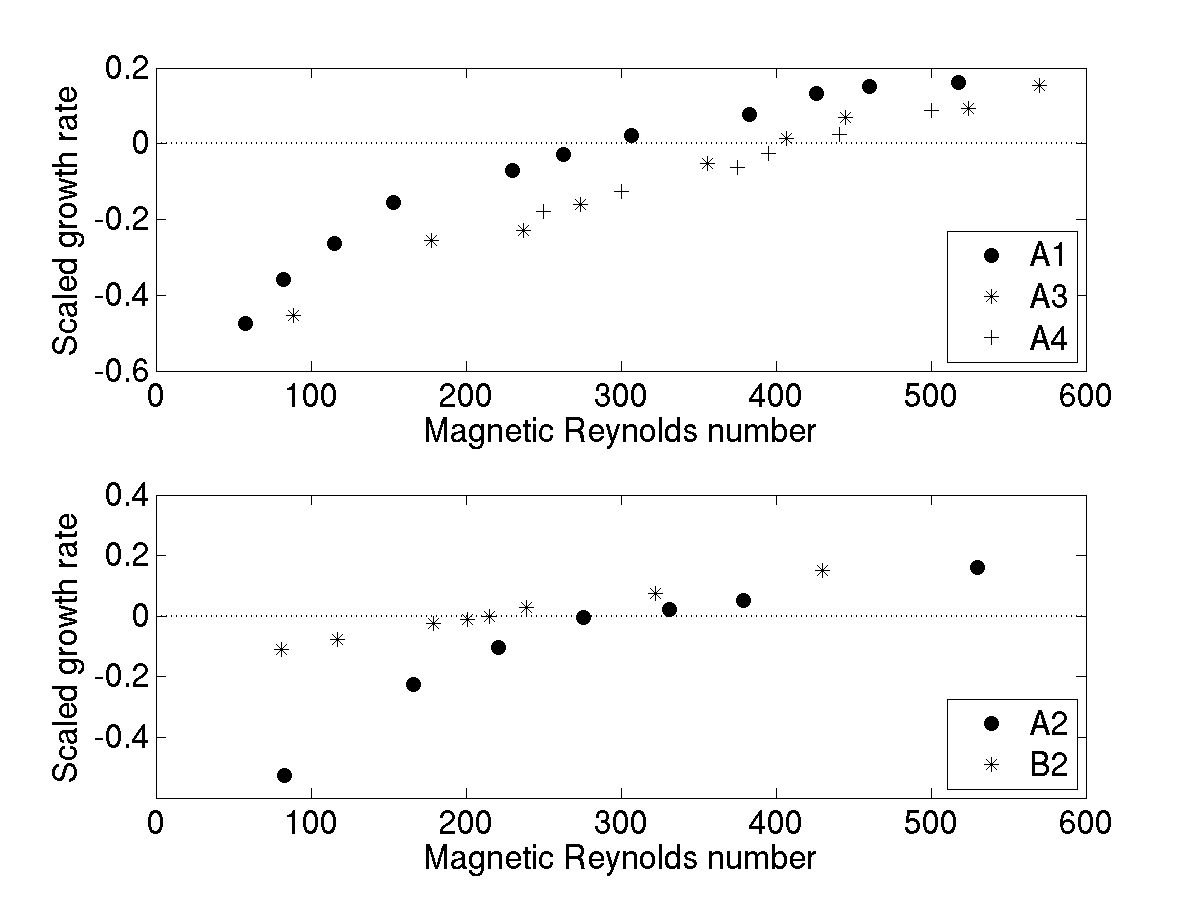}
\caption{Growth rate of the magnetic energy in the kinematic dynamo regime. Top: Cases A1, A3 and A4. Bottom: Cases A2 and B2. In each plot, the growth rate has been scaled by the convective turnover time, $t_{\rm conv}$.} 
\label{fig6}
\end{center}
\end{figure}

\begin{figure}[h]
\begin{center}
\includegraphics[width=9.0cm]{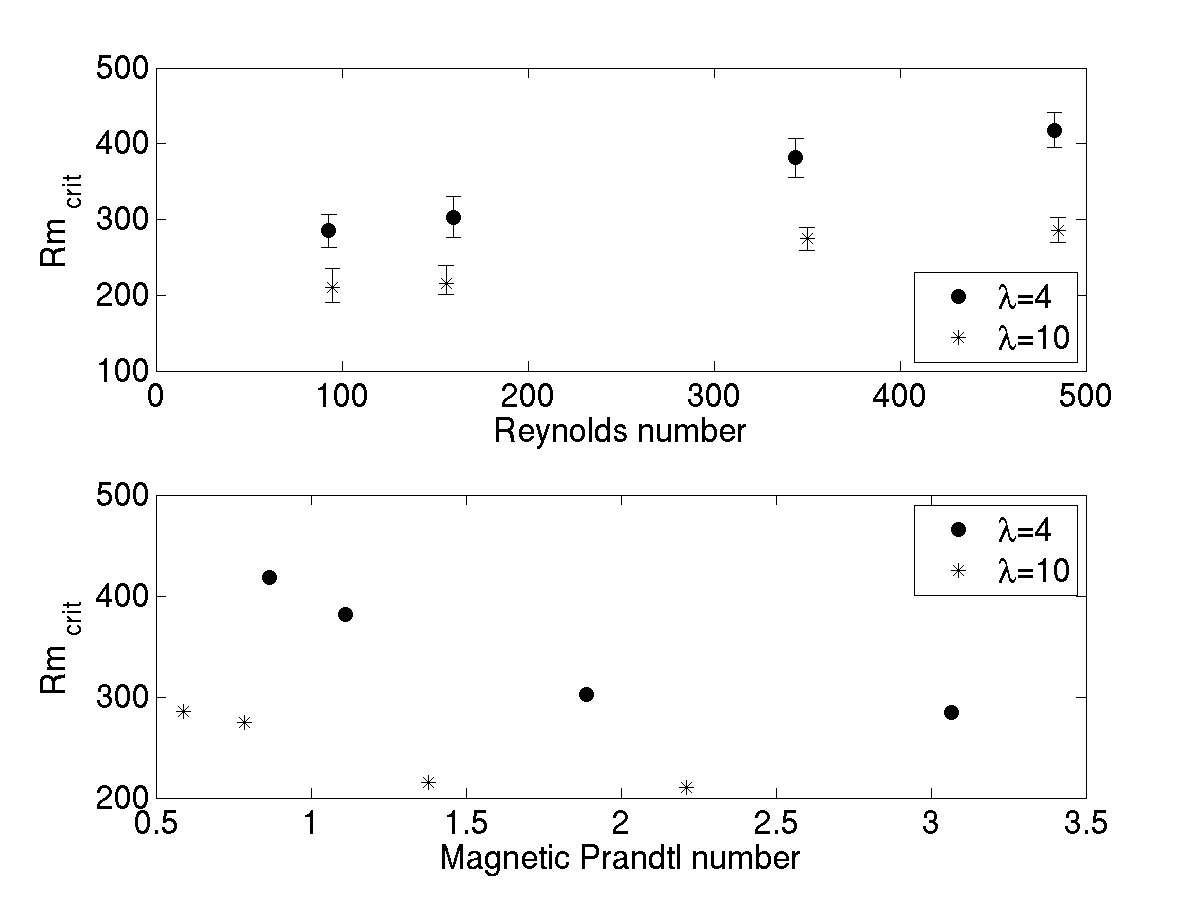}
\caption{Critical magnetic Reynolds number as a function of the Reynolds number (top) and the magnetic Prandtl number (bottom). The circles correspond to the $\lambda=4$ cases, whilst the stars represent the $\lambda=10$ calculations.}
\label{fig7}
\end{center}
\end{figure}

\begin{figure}[ht]
\begin{center}
\includegraphics[width=7.0cm]{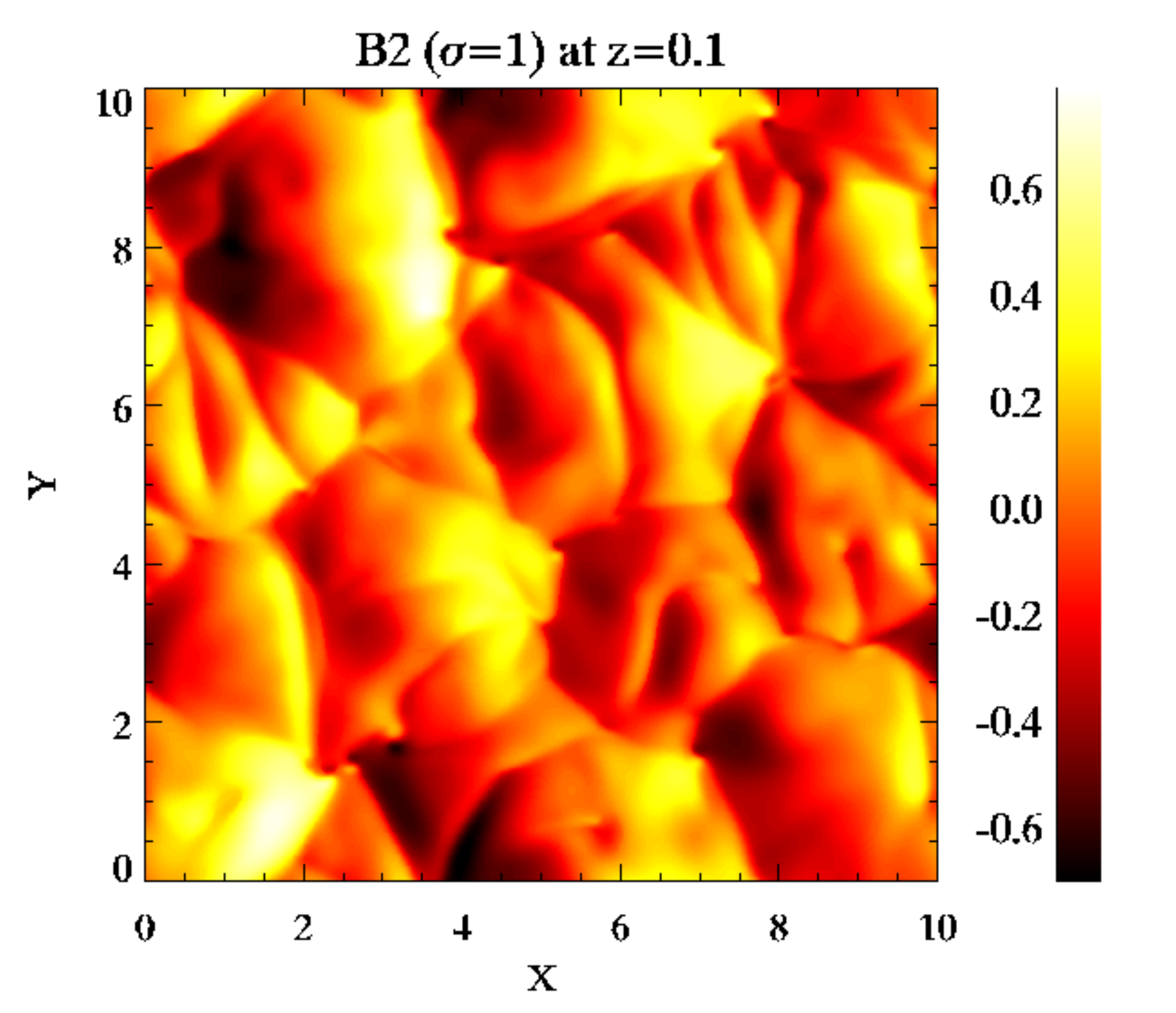}
\includegraphics[width=7.0cm]{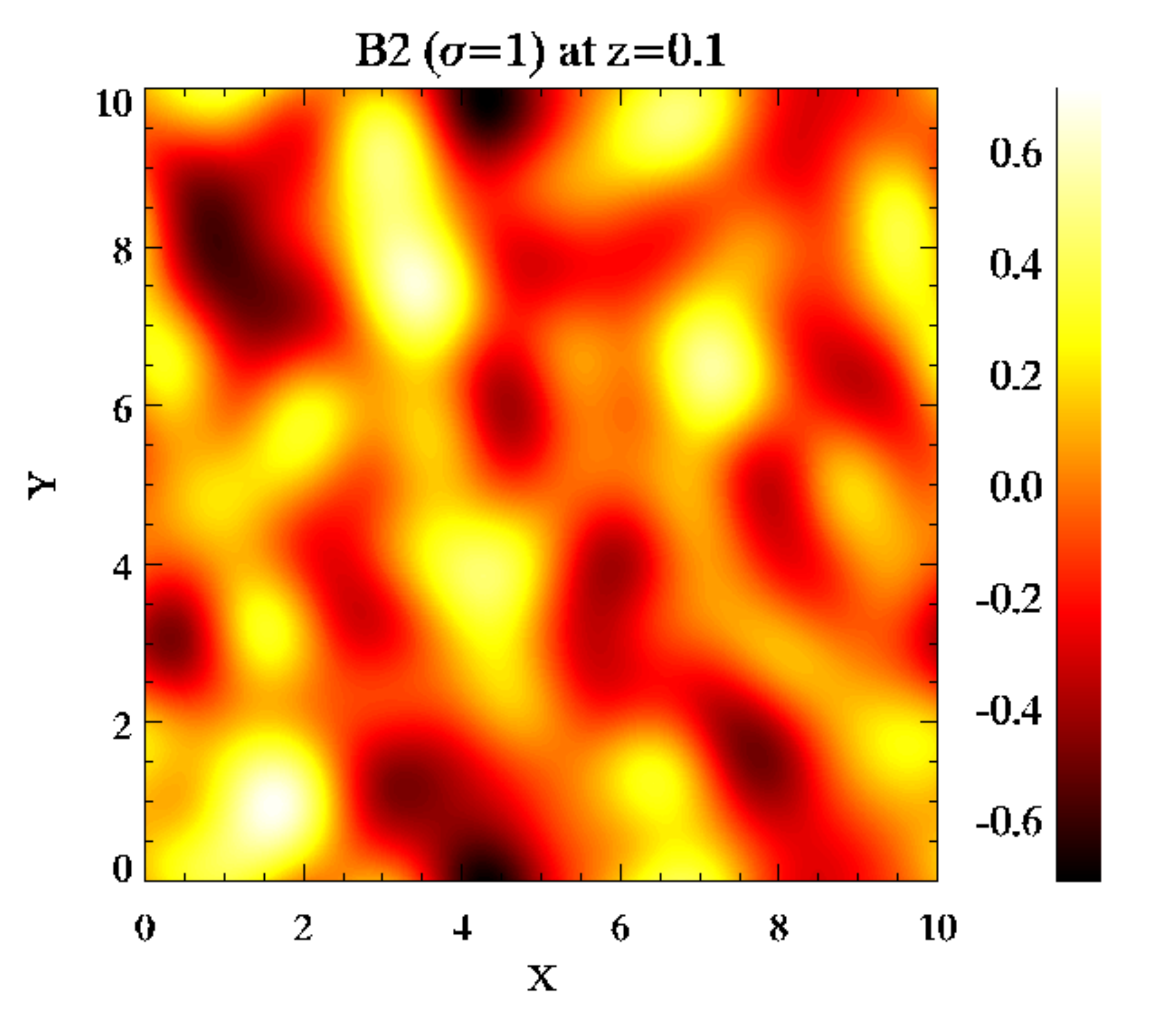} 
\includegraphics[width=7.0cm]{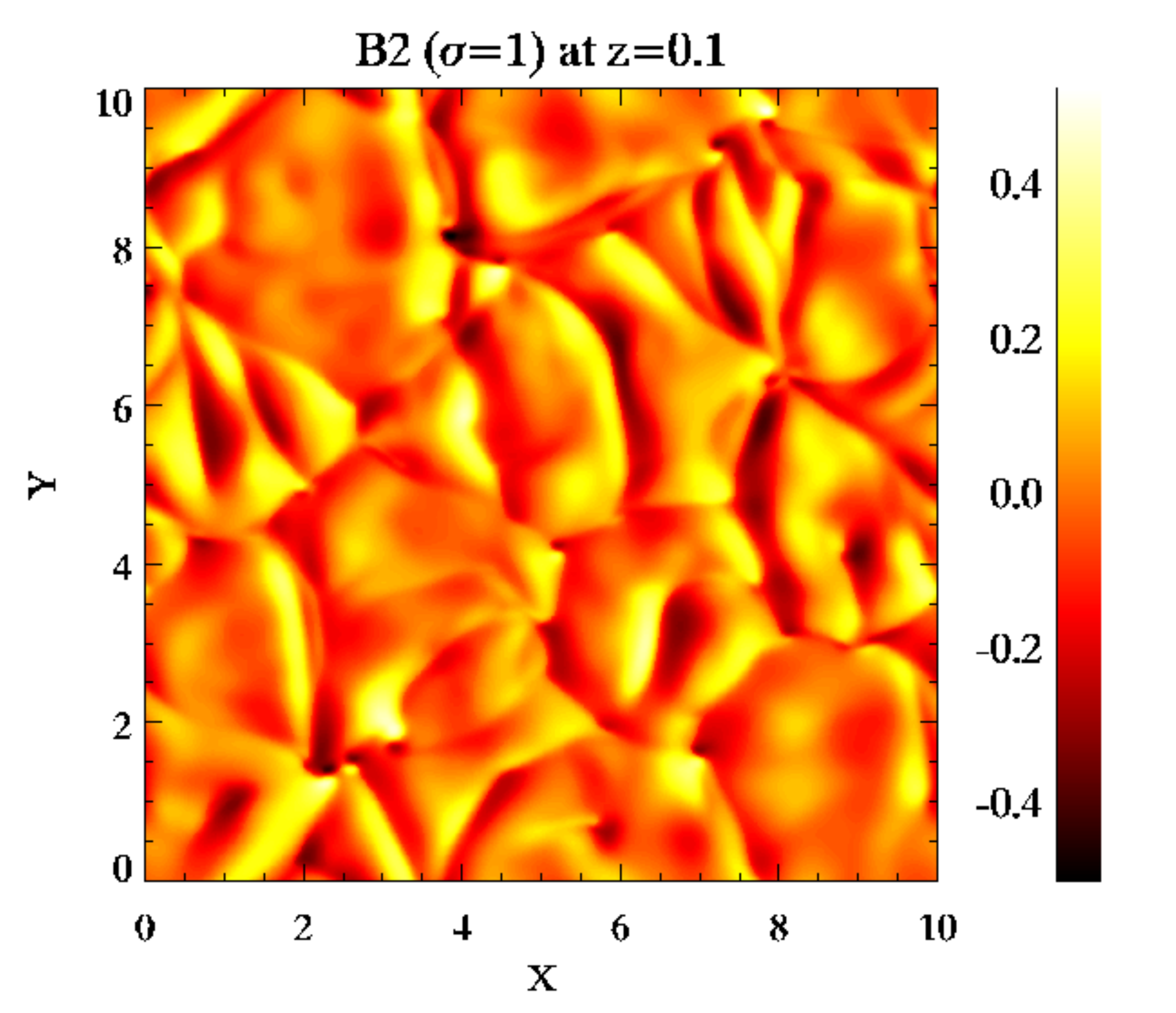}
\caption{Horizontal velocity in the $x$ direction in the horizontal plane $z=0.1$ for the $\lambda=10$, $\sigma=1.0$ case (B2). The upper plot shows the complete velocity field, the middle plot shows the large-scale velocity field (where short wavelengths are removed in Fourier space) and the lower plot shows the small-scale velocity field (where long wavelengths are removed from the flow).} 
\label{fig8}
\end{center}
\end{figure}

\par Figure~\ref{fig6} shows the kinematic growth rates as a function of $Rm$ for cases A1-A4 and B2. Note that the growth rates have been been scaled by the convective turnover time, $t_{\rm conv}$, so as to ensure some degree of comparability between the different flows. Focusing initially upon the upper plot in Figure~\ref{fig6}, which shows the growth rates for three of the $\lambda=4$ cases, we see that the flow with the lowest Reynolds number (A1) is the most efficient dynamo. At comparable values of $Rm$, the growth rates in the higher $Re$ calculations are rather similar, although (in the absence of further calculations at higher Reynolds numbers) it would certainly be premature to suggest that we are approaching an asymptotic regime in which the behaviour is independent of $Re$. The lower plot in Figure~\ref{fig6} shows the $Rm$-dependence of the growth rate for cases A2 and B2. These calculations are comparable in all respects except for the aspect ratio of the domain. At similar values of $Rm$ the growth rates are systematically higher in the larger aspect ratio case, although the most dramatic differences are observed at low $Rm$, where (at comparable parameter values) the magnetic energy decays much more rapidly in the lower aspect ratio case. It is interesting to note that the growth rate appears to have a logarithmic dependence upon $Rm$ in cases A1 and A2 \cite[see also][]{bushbyetal11}, whilst the growth rate curves in the other cases are better approximated by straight lines (at least for the range of values for $Rm$ that has been considered). Logarithmic growth rate curves are not unheard of in the context of turbulent small-scale dynamos \citep[][]{rogachevskiikleeorin97}. However, direct comparison with previous numerical studies of dynamo action in compressible flows \citep[e.g.][]{haugenetal04} is complicated by the fact that, for computational convenience, we hold $Re$ constant whilst varying $Rm$, which implies that the magnetic Prandtl number is not constant. Having said that, it is worth noting that our scaled growth rates at ``high'' $Rm$ are comparable to those found by \citet{haugenetal04} in their study of dynamo action in forced (non-helical) turbulence.

\begin{figure}[t]
\begin{center}
\includegraphics[width=9.0cm]{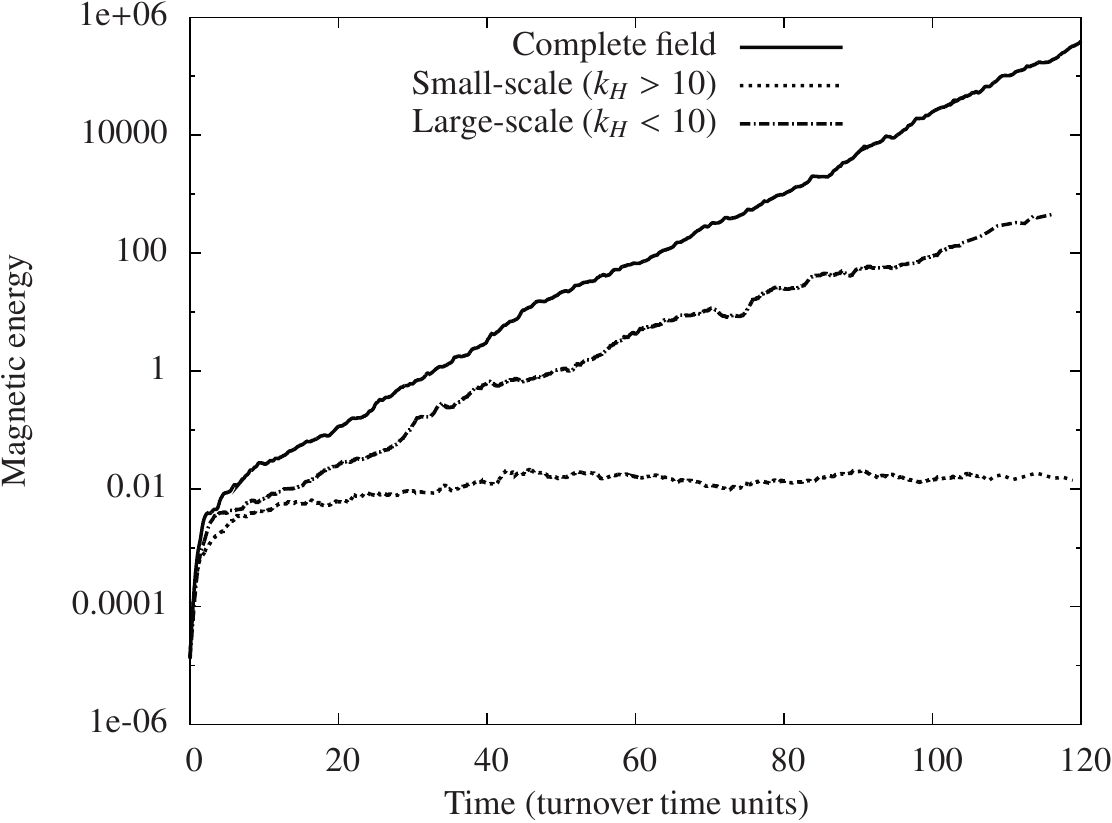}
\caption{Magnetic energy versus time for a kinematic dynamo simulation for the B2 case ($\zeta_0=0.15$). The solid line corresponds to the full dynamo problem in which the induction equation is solved using the complete velocity field. The dotted line corresponds to the dynamo that is driven by the small-scale flow (only modes such that $k_H>10$ are retained in the velocity field that is used to solve the induction equation), whilst the dash-dotted line corresponds to the case in which only the  large-scale flow (where only modes such that $k_H<10$ are retained) has been used.} 
\label{fig9}
\end{center}
\end{figure}

\par Our estimates for $Rm_{\rm crit}$ as a function of the Reynolds number and the magnetic Prandtl number are shown in Figure~\ref{fig7}. The error bars in the upper plot give some indication of the level of precision of these estimates, which are bounded above and below by dynamo calculations which have positive and negative growth rates respectively. The main conclusion that can be drawn from the upper plot is that (at comparable values of $Re$) the critical magnetic Reynolds number is always smaller in the larger aspect ratio calculations. This indicates that it is always easier to drive a dynamo in the larger domain, where mesogranulation is present. \citet{bushbyetal12} came to a similar conclusion although they only considered a single flow -- the key point here is that this conclusion seems to be independent of the Reynolds number, which suggests that it is a robust feature of this system. The lower plot in Figure~\ref{fig7} indicates that dynamo action is always easier to excite at higher values of the magnetic Prandtl number, with systematically higher values of $Rm_{\rm crit}$ in the lower $Pm$ cases. The relative inefficiency of turbulent dynamos at low $Pm$ is well known \citep[][]{boldyrevcattaneo04,schekochihinetal05}, so this result is unsurprising. It is worth noting here that estimates suggest that $Pm \approx 10^{-6}$ in the solar photosphere \citep[see, e.g.,][]{ossendrijver03}. Due to computational constraints, the smallest value of the magnetic Prandtl number for which we have found a dynamo is $Pm \approx 0.5$, so we are clearly not in a position to say anything definitive about $Rm_{\rm crit}$ for the convection in the near-surface layers of the Sun. However, the local magnetic Reynolds number in this part of the Sun is very large \citep[$Rm\approx 10^6$, see][]{ossendrijver03} so it would be surprising if a local dynamo was not operating in this region.  

\subsection{Filtered velocity field}

These kinematic dynamo results show that the dynamo is more efficient in a larger domain. This suggests that the mesogranular component of the flow may be playing a crucial role in promoting dynamo action in the $\lambda=10$ cases. After the mesogranular peak, the kinetic energy spectra that are shown in Figure~\ref{fig3} decrease monotonically with increasing $k_H$. The fact that there is no secondary peak corresponding to a granular scale means that there is no clear separation in scales between the motions at large and small scales. Nevertheless, it is of interest to try to determine the ways in which components of the flow at different spatial scales influence the dynamo.  In order to investigate this issue, we filter the velocity field in Fourier space. At each time-step within the numerical scheme, the velocity field is Fourier-transformed in the two horizontal directions. The filtered velocity field is obtained by multiplying the depth-dependent Fourier coefficients by the following mask function
\begin{equation}
\label{filter}
\hat{f}(k_x,k_y) = \left\{
    \begin{array}{ll}
        1 & \mbox{if } \sqrt{k_x^2+k_y^2} \le k_c \\
        0 & \mbox{otherwise} \ ,
    \end{array}
\right.
\end{equation}
where $k_c$ is an arbitrary cut-off wavenumber. By suppressing the energy contained in wavenumbers larger than $k_c$, we isolate the large-scale component of the velocity field. Similarly, one can extract the small-scale component of the velocity field by ignoring wavenumbers smaller than $k_c$. In this approach, the full velocity field is then used in all of the governing equations except in the induction equation~(\ref{eq:induction}), where the filtered velocity field is used to determine the evolution of the magnetic field. A similar filtering method has already been used by \cite{hughes2013} to study the effect of shear on convectively-driven dynamos and by \cite{tobias2008} to study fast dynamos. This simple filtering process does not preserve coherent structures in physical space, since the mask function only depends on the wavenumber and not on the phase of the Fourier coefficients. However, despite its limitations, this approach should give us some insights into the role of mesogranulation. In the following, we use the same parameters as for simulation B2 in Table~\ref{table1}. We choose a cut-off wavenumber of $k_c=10$ which lies to the right of the mesogranular peak in the kinetic energy spectrum that is shown in Figure~\ref{fig3}. 

\begin{figure*}[th]
\begin{center}
\includegraphics[width=8.5cm]{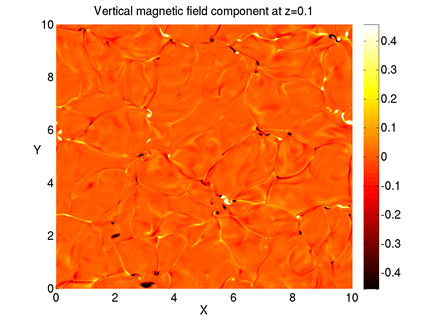}
\includegraphics[width=8.5cm]{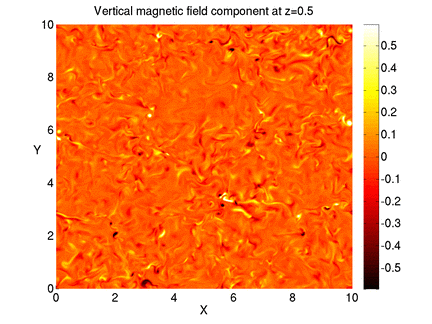}
\includegraphics[width=8.5cm]{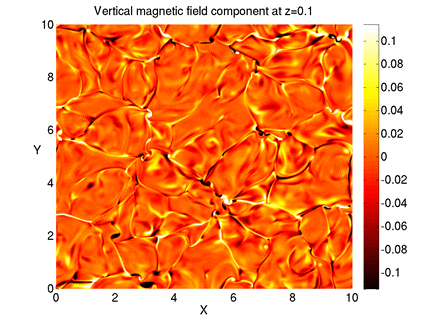}
\includegraphics[width=8.5cm]{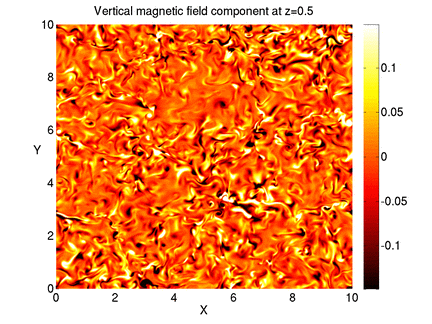}
\includegraphics[width=8.5cm]{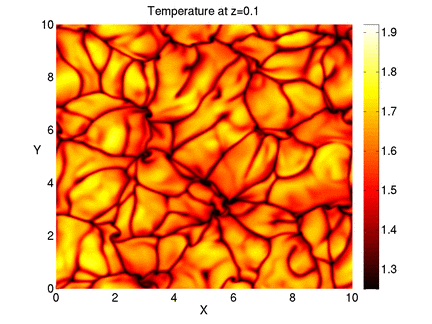}
\includegraphics[width=8.5cm]{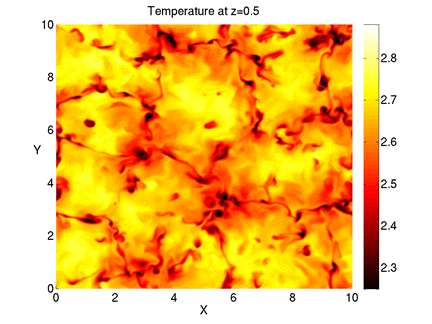}
\caption{Snapshot of a statistically-steady nonlinear dynamo calculation (corresponding to case B4, with $\zeta_0=0.15$ -- see Table~\ref{table2}). Top: The vertical magnetic field distribution in the horizontal planes defined by $z=0.1$ (Left) and $z=0.5$ (Right), where the colour table has been saturated at $B_z=\pm B_{\rm eq}$. Middle: As above, but here the contour levels have been saturated at $\pm 0.25 B_{\rm eq}$ to highlight some of the small-scale structure in the field distribution. Bottom: The corresponding temperature distributions for $z=0.1$ (Left) and $z=0.5$ (Right).}
\label{fig10}
\end{center}
\end{figure*} 

\par The effects of this filtering process are illustrated in Figure~\ref{fig8}, which shows snapshots of the full and filtered velocity fields for this simulation, whilst Figure~\ref{fig9} shows the time-evolution of the magnetic energy for the corresponding kinematic dynamo calculations (for $\zeta_0=0.15$). Although the magnetic energy grows at a slightly slower rate for the large-scale flow than it does in the unfiltered case, the dynamo efficiencies are fairly comparable. However, the small-scale flow is only a marginal kinematic dynamo for these parameter values. This disparity between the dynamo growth rates for the large-scale and the small-scale flows cannot be attributed to there being more kinetic energy in the large-scale motions. In fact the rms velocity associated with the large-scale flow is approximately the same as for the small-scale flow ($U_{rms}\approx 0.25$), which implies that the effective magnetic Reynolds number is approximately the same for both filtered flows. Therefore these results do seem to confirm the notion that the flows at scales that are comparable to the mesogranular scale are more effective than the small-scale flows when it comes to driving a kinematic dynamo in this particular system. 

\subsection{Nonlinear results and comparison to observations}\label{subsec:nondynamo}

Having established the conditions under which this system can excite a kinematic dynamo, we now return to the full nonlinear problem in which the magnetic field is allowed to exert a dynamical influence upon the flow. One particular nonlinear calculation is illustrated in Figure~\ref{fig10}, which shows a snapshot of the temperature and magnetic field distribution in two different horizontal planes after the dynamo has evolved into a statistically-steady state. This dynamo is based upon case $B4$, which is the highest Reynolds number $\lambda=10$ simulation that has been considered in this paper, whilst a value of $\zeta_0=0.15$ was chosen when the initial seed magnetic field was introduced into the flow. In this nonlinear state, the rms velocity ($U_{\rm rms}=0.390$) is approximately $5\%$ smaller than that of the original hydrodynamic flow. This implies that the corresponding Reynolds numbers are $Re=458$ and $Rm=475$ (with $Pm=1.04$). The near-surface temperature distribution that is shown in Figure~\ref{fig10} is qualitatively similar to the one illustrated in Figure~\ref{fig1}, whilst the mid-layer temperature distribution is again indicative of the presence of mesogranulation. The fact that the temperature distribution is rather similar to that of the corresponding hydrodynamic flow suggests that the magnetic field only has a local (rather than global) effect upon the convective dynamics. The contour levels in the corresponding magnetic field plots are related to the time-averaged equipartition field strength, $\displaystyle{B_{\rm eq}(z) = \overline{0.5\rho|\vec{u}|^2}}$, at that particular depth (where the overbar corresponds to a horizontal spatial average). In the upper plots in Figure~\ref{fig10}, the colour table has been saturated at $\pm B_{\rm eq}$, whilst the colour table in the middle plots has been saturated at  $\pm 0.25 B_{\rm eq}$. Comparing the near-surface magnetic field (in particular) with the mid-layer temperature distribution, it is evident that the stronger vertical magnetic fields tend to be associated with the mesogranular boundaries, which correspond to the locations of the strongest convective downflows. This is, therefore, at least qualitatively consistent with observations of quiet Sun magnetic fields. Near the upper surface, the intergranular lanes tend to be associated with weaker magnetic fields, whilst very weak fields can also be seen in some of the granular interiors. These presumably correspond to loops of field that are advected into the surface layers by the convective upflows. As illustrated in Figure~\ref{fig10}, the weak magnetic field is almost ``space filling" at the mid-plane of the domain, where a complex, mixed-polarity field distribution can be observed.

\par
\begin{figure}[t]
\begin{center}
\includegraphics[width=9.0cm]{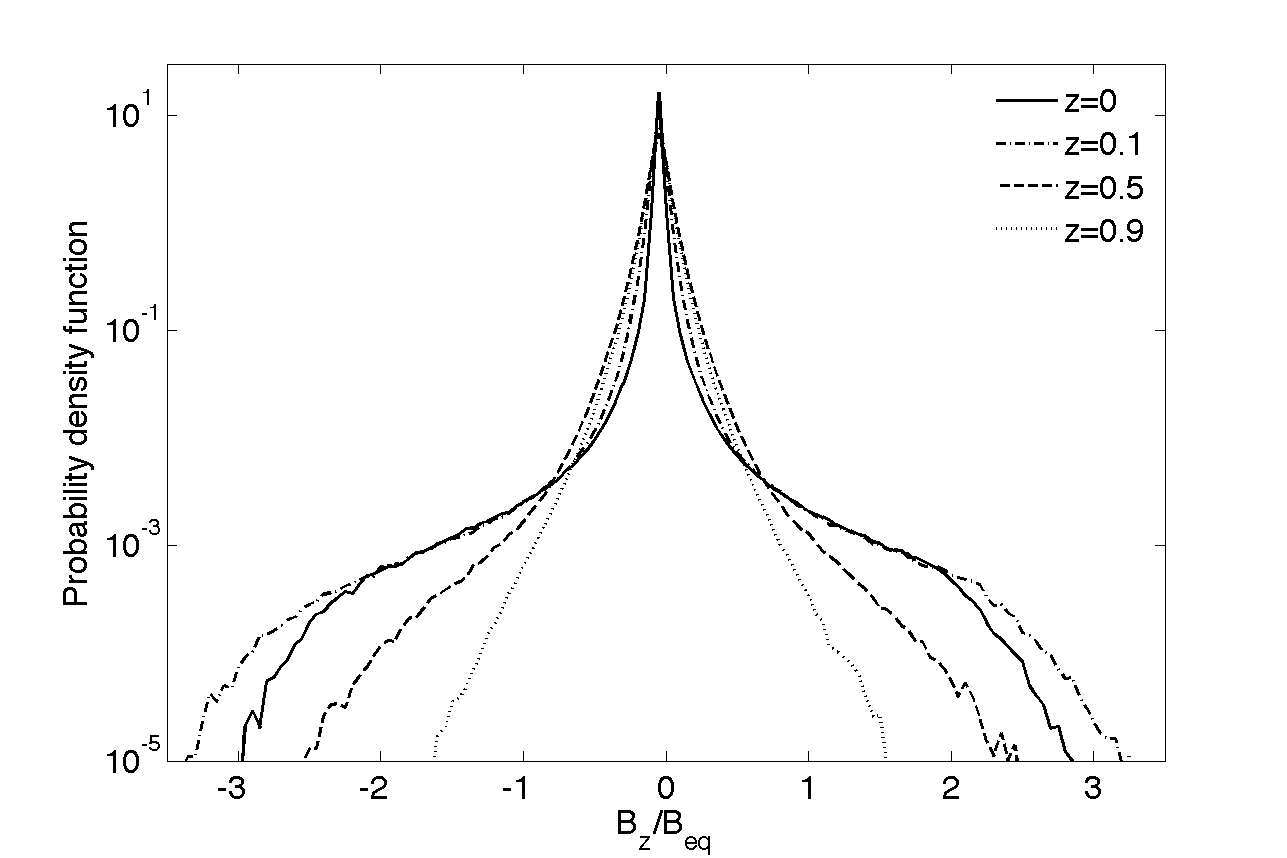}
\caption{Time-averaged probability density function for the vertical component of the magnetic field at 4 different depths for the statistically-steady, nonlinear dynamo calculation that is illustrated in Figure~\ref{fig10}. The different curves correspond to $z=0$ (solid line), $z=0.1$ (dash-dotted line), $z=0.5$ (dashed line) and $z=0.9$ (dotted line). At each depth, the magnetic field has been normalised by the local equipartition field strength, $B_{\rm eq}$.} 
\label{fig11}
\end{center}
\end{figure}

 Still focusing upon the calculation that is illustrated in Figure~\ref{fig10}, Figure~\ref{fig11} shows the corresponding (time-averaged) probability density functions (PDFs) for the vertical component of the magnetic field at four different depths within the domain. At each depth, the magnetic field strength has again been normalised by the local equipartition magnetic field strength. Near the surface, the PDFs take the form of stretched exponential distributions, with the tail extending well into the ``super-equipartition'' range (with a peak field that is roughly three times the equipartition value). At lower values of $Re$, similar PDFs can be found in nonlinear dynamo calculations at comparable values of $Rm$ \citep[][]{bushbyetal12} so this tendency to produce super-equipartition fields does not appear to be strongly dependent upon either $Re$ or $Pm$ (at least for computationally accessible values of these parameters). To relate these peak field strengths to solar observations, we need to bear in mind that $B_{\rm eq}\approx 400$G near the solar surface \citep[][]{gallowayetal77}. The strongest vertical magnetic fields in these simulations would therefore correspond to the kG-strength fields that are observed in the quiet Sun. As discussed in the Introduction, the mechanism for the formation of these strong fields is related to the well-known convective collapse instability, and the resultant concentrations of vertical magnetic flux are partially evacuated. For this calculation, the minimum surface density within one of these magnetic regions is (typically) of order of $30\%$ of the mean surface density, although it can be substantially smaller than this. Looking again at Figure~\ref{fig11}, it is clear that the PDFs at $z=0$ and $z=0.1$ are rather similar, although slightly stronger peak fields (relative to $B_{\rm eq}$) are found at $z=0.1$. Even at $z=0.5$ the peak field strength can significantly exceed the equipartition value. In the deeper layers, the PDF is well-approximated by a standard exponential distribution in which the peak field strength is comparable to $B_{\rm eq}$. Note that all of these PDFs are almost symmetric about $B_z/B_{\rm eq}=0$. This implies a mixed polarity field distribution at all depths with no preference for either polarity. This result reflects the absence of any imposed magnetic fields in these calculations.

\par When comparing these calculations to observations of magnetic fields in the quiet Sun, it is also of interest to measure the mean (unsigned) vertical flux across the surface of the domain, $\overline{|B_z|}$. In the study of \citet{litesetal08}, they measured the mean value of the modulus of the line of sight component of $\vec{B}$ to be approximately $11G$, which is slightly less than $3\%$ of the local equipartition field strength. Defining $\overline{|B_z|}_{\rm ,0}$ to be the surface value of the mean of  $|B_z|$ and $B_{\rm eq, 0}$ to be the corresponding equipartition field strength, this simulation gives a time-averaged value of $\overline{|B_z|}_{\rm ,0}/B_{\rm eq, 0} \approx 0.027$, which is remarkably similar to the value quoted by  \citet{litesetal08}. Given the idealised nature of this model, and the fact that we are comparing numerical results to lower resolution observations, we should not read too much into the fact that we have such close agreement in this case. However, it is clearly encouraging that the mean value of $|B_z|$ is of the correct order of magnitude to be consistent with observations. For completeness, it is worth noting that the quantity $\overline{|B_z|}/B_{\rm eq}$ is actually an increasing function of depth until just below the mid-layer (reaching a peak value of approximately $0.075$) before decreasing again with depth in the lower part of the domain. The time-average of the ratio of the total magnetic energy ($ME$) to the total kinetic energy ($KE$) is of the same order of magnitude as $\overline{|B_z|}/B_{\rm eq}$ (here $ME/KE=0.030$).

\begin{table}
\begin{center}
\caption{.\label{table2}Summary of the properties of the statistically-steady nonlinear dynamo calculations.} 
\begin{tabular}{cccccccc} 
\hline\hline
Run & $\zeta_0$  & $U_{\rm rms}$ & $Re$ & $Rm$ & $Pm$ & $\displaystyle{\frac{\overline{|B_z|_{\rm ,0}}}{B_{\rm eq, 0}}}$ & $\displaystyle{\frac{ME}{KE}}$ \\ \hline
A1 & $0.111$ & $0.304$ & $89$ & $500$ & $5.62$  & $0.031$ & $0.076$ \\
A2 & $0.125$ & $0.337$ & $149$ & $492$ & $3.30$ & $0.024$ & $0.047$ \\
A3 & $0.136$ & $0.385$ & $339$ & $517$ & $1.53$ & $0.020$ & $0.025$ \\
A4 & $0.15$ & $0.409$ & $479$ & $498$ & $1.04$ & $0.012$ & $0.008$ \\
\hline
B1 & $0.15$ & $0.303$ & $89$ & $369$ & $4.15$ & $0.031$ & $0.059$ \\
B2 & $0.15$ & $0.336$ & $148$ & $409$ & $2.76$ & $0.029$ & $0.048$ \\
B3 & $0.15$ & $0.378$ & $333$ & $460$ & $1.38$ & $0.028$ & $0.037$ \\
B4 & $0.15$ & $0.390$ & $458$ & $475$ & $1.04$ & $0.027$ & $0.030$ \\
\hline
\end{tabular}
\tablefoot{The convective flows correspond to those given in Table~\ref{table1}. As before,  $U_{\rm rms}$ corresponds to the time-averaged rms velocity, and both $Re$ and $Rm$ have been derived from this. The quantity $\overline{|B_z|}_{\rm ,0}$ corresponds to the horizontally-averaged value of the modulus of $B_z$ at the surface, $B_{\rm eq, 0}$ is the equipartition field strength at $z=0$, whilst $KE$ and $ME$ represent the total kinetic and magnetic energies respectively. All of these quantities have also been time-averaged.}
\end{center}
\end{table} 

\par One of the main aims of this study was to assess the $Re$ dependence of these convectively-driven dynamos. We therefore carried out a range of nonlinear calculations (which are summarised in Table~\ref{table2}). The presence of partially-evacuated regions (which, for various reasons, reduces the critical time-step for the stability of the numerical scheme) greatly increases the computational demands relative to the corresponding kinematic calculations. We therefore carried out a more restricted parameter survey for the full nonlinear problem. However, it is still possible to identify clear trends in the parameter-dependence of the statistics that were measured. In the $\lambda=4$ calculations, we chose values of $\zeta_0$ that produced dynamos with similar values of $Rm$. In that case, both $\overline{|B_z|}_{\rm ,0}/B_{\rm eq, 0}$ and $ME/KE$ increase with increasing $Pm$ (or, equivalently, decreasing $Re$). So, at comparable values of $Rm$, the nonlinear efficiency of the dynamo is greatest in the higher $Pm$/lower $Re$ cases. This can be attributed to the fact that $Rm_{\rm crit}$ is lower in the higher $Pm$ cases, which implies that the dynamos in these cases are further above onset than those in the lower $Pm$ regime. For computational convenience, we adopted a slightly different approach in the more computationally expensive $\lambda=10$ cases in that we fixed the input parameter $\zeta_0=0.15$ (as opposed to adjusting it to attain a dynamo at a certain value of $Rm$). Although $\overline{|B_z|}_{\rm ,0}/B_{\rm eq, 0}\approx 0.03$ in all of these cases, there is again a tendency for larger $ME/KE$ ratios in the higher $Pm$ cases. The two dynamo simulations corresponding to A4 and B4 are comparable in every respect except for the differing aspect ratios. The higher level of efficiency in the larger aspect ratio B4 case is again simply a consequence of the fact that it is further from onset than the corresponding A4 calculation. 
  
\section{Summary and discussion}\label{sec:summary}

Motivated by observations of convection and small-scale magnetic fields in regions of quiet Sun, we have investigated the dynamo properties of compressible convection in a large aspect ratio domain. From a purely hydrodynamic perspective, the first key result from this study is that mesogranulation seems to be a robust feature of models of this type, although a wide domain is needed in order to accommodate these large-scale features. Furthermore, over the range of computationally accessible values of the Reynolds number, we have found that the peak at the mesogranular scale in the kinetic energy spectrum is more pronounced in the higher $Re$ cases. All of the convective flows that were considered in this paper are capable of driving a dynamo, provided that the magnetic Reynolds number exceeds some critical value. This critical value tends to increase with increasing values of $Re$. However $Rm_{\rm crit}$ is always smaller in the $\lambda=10$ cases than it is in comparable calculations in the $\lambda=4$ domain, and for a given value of $Rm>Rm_{\rm crit}$ the kinematic growth rates are always higher in the $\lambda=10$ case. In addition to this aspect ratio dependence, our analysis of the dynamo properties of velocity fields that have been filtered in Fourier space also seems to confirm the suggestion of  \citet{bushbyetal12} that mesogranulation is beneficial for dynamo action. In the nonlinear regime, near-surface vertical magnetic flux concentrations tend to accumulate preferentially at the boundaries of mesogranules, where peak field strengths can significantly exceed the equipartition field strength (an observation that is consistent with the kG-strength fields that are observed in the quiet Sun). The horizontally-averaged values of $|B_z|$ at the upper surface of the domain are also of the correct order of magnitude to be comparable to observed values. In general, the nonlinear efficiency of the dynamo (the ratio of the total magnetic energy to the total kinetic energy) increases with increasing values of the magnetic Prandtl number, although the magnetic energy never exceeds $10\%$ of the kinetic energy in the parameter regimes that are investigated in this paper. 

\par It should be stressed again that it is not possible to carry out simulations of this type in realistic solar-like parameter regimes, so some care is needed when relating results to observations. However, our findings do tend to favour the existence of a genuine mesogranular scale of motion in the near-surface convective flows. As we have noted, the $Re$-dependence of the vertical vorticity distribution provides some support for the hypothesis that there is a relationship between these vortices and the mesogranulation, although this causal connection is still rather speculative. Nevertheless, the fact that mesogranulation appears to be absent in rotating turbulent convection \citep[in which vortices tend to align themselves with the rotation axis, see, e.g.,][]{brummelletal98}, provides some support for the idea that the vortices may be playing an important role in this context. Another issue that needs to addressed here is the question of why some simulations naturally produce mesogranulation whilst others do not. Some numerical schemes rely upon artificial viscosities, which can have a strong dissipative effect at small scales. If this strongly influences the vorticity distribution, which is structured on small scales, this may explain why calculations of this type do not tend to produce mesogranulation with a well-defined spatial scale. However, this is a hypothesis that needs to be tested, so this is an obvious area for future work.

\par Having established that the convective flows in this idealised model can produce dynamo-generated magnetic fields that compare favourably to observations, the next step will be to relax some of the simplifying assumptions that have been made. Whilst the assumption of periodic boundary conditions in the horizontal direction is perfectly reasonably for a local model of this type, we should not rule out the possibility that the upper and lower boundary conditions may be playing an important role in determining the behaviour of the system. Some calculations assume open upper and lower boundaries, and we have carried out some test calculations for an analogous system. Although results are still at a rather preliminary stage, they do seem to indicate the presence of mesogranulation, which suggests that this phenomenon is not strongly-dependent upon the choice of hydrodynamic boundary conditions. This system also allows us to relax some of the constraints due to the magnetic boundary conditions. An initial dynamo calculation suggests that the near-surface vertical magnetic field distribution is qualitatively similar to that observed in the more idealised calculations that are described in this paper (despite the presence of additional horizontal magnetic fields at the same level). This preliminary calculation therefore indicates that the operation of the dynamo may not be strongly-dependent upon the choice of magnetic field boundary conditions, although further work needs to be done in order to test this hypothesis. Finally, it is worth noting that the convective domain that is described in this paper is relatively shallow in terms of pressure scale heights. Increasing the number of scale heights in the system may have an effect upon the efficiency of the dynamo, but again further study is needed in order to investigate this issue. 

\begin{acknowledgements}
All calculations were carried out on the UK MHD Consortium computing facilities at St Andrews and Warwick. B. Favier thanks the Cambridge Newton Trust for financial support.
\end{acknowledgements}

\bibliographystyle{aa} 
\bibliography{biblio}

\begin{thebibliography}{58}
\expandafter\ifx\csname natexlab\endcsname\relax\def\natexlab#1{#1}\fi

\bibitem[{{Abbett}(2007)}]{abbett07}
{Abbett}, W.~P. 2007, \apj, 665, 1469

\bibitem[{{Boldyrev} \& {Cattaneo}(2004)}]{boldyrevcattaneo04}
{Boldyrev}, S. \& {Cattaneo}, F. 2004, Physical Review Letters, 92, 144501

\bibitem[{{Brummell} {et~al.}(2010){Brummell}, {Tobias}, \&
  {Cattaneo}}]{brummelletal10}
{Brummell}, N., {Tobias}, S., \& {Cattaneo}, F. 2010, Geophysical and
  Astrophysical Fluid Dynamics, 104, 565

\bibitem[{{Brummell} {et~al.}(1998){Brummell}, {Hurlburt}, \&
  {Toomre}}]{brummelletal98}
{Brummell}, N.~H., {Hurlburt}, N.~E., \& {Toomre}, J. 1998, \apj, 493, 955

\bibitem[{{Buehler} {et~al.}(2013){Buehler}, {Lagg}, \&
  {Solanki}}]{buehleretal13}
{Buehler}, D., {Lagg}, A., \& {Solanki}, S.~K. 2013, \aap, 555, A33

\bibitem[{{Bushby} {et~al.}(2012){Bushby}, {Favier}, {Proctor}, \&
  {Weiss}}]{bushbyetal12}
{Bushby}, P.~J., {Favier}, B., {Proctor}, M.~R.~E., \& {Weiss}, N.~O. 2012,
  Geophysical and Astrophysical Fluid Dynamics, 106, 508

\bibitem[{{Bushby} \& {Houghton}(2005)}]{bushbyhoughton05}
{Bushby}, P.~J. \& {Houghton}, S.~M. 2005, \mnras, 362, 313

\bibitem[{{Bushby} {et~al.}(2008){Bushby}, {Houghton}, {Proctor}, \&
  {Weiss}}]{bushbyetal08}
{Bushby}, P.~J., {Houghton}, S.~M., {Proctor}, M.~R.~E., \& {Weiss}, N.~O.
  2008, \mnras, 387, 698

\bibitem[{{Bushby} {et~al.}(2010){Bushby}, {Proctor}, \&
  {Weiss}}]{bushbyetal10}
{Bushby}, P.~J., {Proctor}, M.~R.~E., \& {Weiss}, N.~O. 2010, in Astronomical
  Society of the Pacific Conference Series, Vol. 429, Numerical Modeling of
  Space Plasma Flows, Astronum-2009, ed. N.~V. {Pogorelov}, E.~{Audit}, \&
  G.~P. {Zank}, 181

\bibitem[{{Bushby} {et~al.}(2011){Bushby}, {Proctor}, \&
  {Weiss}}]{bushbyetal11}
{Bushby}, P.~J., {Proctor}, M.~R.~E., \& {Weiss}, N.~O. 2011, in IAU Symposium,
  Vol. 271, IAU Symposium, ed. N.~H. {Brummell}, A.~S. {Brun}, M.~S. {Miesch},
  \& Y.~{Ponty}, 197--204

\bibitem[{{Cattaneo}(1999)}]{cattaneo99}
{Cattaneo}, F. 1999, \apjl, 515, L39

\bibitem[{{Cattaneo} {et~al.}(2001){Cattaneo}, {Lenz}, \&
  {Weiss}}]{cattaneoetal01}
{Cattaneo}, F., {Lenz}, D., \& {Weiss}, N. 2001, \apjl, 563, L91

\bibitem[{{Danilovic} {et~al.}(2010){Danilovic}, {Sch{\"u}ssler}, \&
  {Solanki}}]{danilovicetal10}
{Danilovic}, S., {Sch{\"u}ssler}, M., \& {Solanki}, S.~K. 2010, \aap, 513, A1

\bibitem[{{de Wijn} {et~al.}(2008){de Wijn}, {Lites}, {Berger}, {Frank},
  {Tarbell}, \& {Ishikawa}}]{wijnetal08}
{de Wijn}, A.~G., {Lites}, B.~W., {Berger}, T.~E., {et~al.} 2008, \apj, 684,
  1469

\bibitem[{{de Wijn} {et~al.}(2005){de Wijn}, {Rutten}, {Haverkamp}, \&
  {S{\"u}tterlin}}]{wijnetal05}
{de Wijn}, A.~G., {Rutten}, R.~J., {Haverkamp}, E.~M.~W.~P., \&
  {S{\"u}tterlin}, P. 2005, \aap, 441, 1183

\bibitem[{{Dom{\'{\i}}nguez Cerde{\~n}a}(2003)}]{dominguez03}
{Dom{\'{\i}}nguez Cerde{\~n}a}, I. 2003, \aap, 412, L65

\bibitem[{{Dom{\'{\i}}nguez Cerde{\~n}a}
  {et~al.}(2006{\natexlab{a}}){Dom{\'{\i}}nguez Cerde{\~n}a}, {S{\'a}nchez
  Almeida}, \& {Kneer}}]{dominguezetal06b}
{Dom{\'{\i}}nguez Cerde{\~n}a}, I., {S{\'a}nchez Almeida}, J., \& {Kneer}, F.
  2006{\natexlab{a}}, \apj, 646, 1421

\bibitem[{{Dom{\'{\i}}nguez Cerde{\~n}a}
  {et~al.}(2006{\natexlab{b}}){Dom{\'{\i}}nguez Cerde{\~n}a}, {S{\'a}nchez
  Almeida}, \& {Kneer}}]{dominguezetal06a}
{Dom{\'{\i}}nguez Cerde{\~n}a}, I., {S{\'a}nchez Almeida}, J., \& {Kneer}, F.
  2006{\natexlab{b}}, \apj, 636, 496

\bibitem[{{Galloway} {et~al.}(1977){Galloway}, {Proctor}, \&
  {Weiss}}]{gallowayetal77}
{Galloway}, D.~J., {Proctor}, M.~R.~E., \& {Weiss}, N.~O. 1977, \nat, 266, 686

\bibitem[{{Grossmann-Doerth} {et~al.}(1998){Grossmann-Doerth}, {Schuessler}, \&
  {Steiner}}]{grossmannetal98}
{Grossmann-Doerth}, U., {Schuessler}, M., \& {Steiner}, O. 1998, \aap, 337, 928

\bibitem[{{Haugen} {et~al.}(2004){Haugen}, {Brandenburg}, \&
  {Dobler}}]{haugenetal04}
{Haugen}, N.~E., {Brandenburg}, A., \& {Dobler}, W. 2004, \pre, 70, 016308

\bibitem[{{Hughes} \& {Proctor}(2013)}]{hughes2013}
{Hughes}, D.~W. \& {Proctor}, M.~R.~E. 2013, Journal of Fluid Mechanics, 717,
  395

\bibitem[{{Ishikawa} \& {Tsuneta}(2011)}]{ishikawatsuneta11}
{Ishikawa}, R. \& {Tsuneta}, S. 2011, \apj, 735, 74

\bibitem[{{K{\"a}pyl{\"a}} {et~al.}(2008){K{\"a}pyl{\"a}}, {Korpi}, \&
  {Brandenburg}}]{kapylaetal08}
{K{\"a}pyl{\"a}}, P.~J., {Korpi}, M.~J., \& {Brandenburg}, A. 2008, \aap, 491,
  353

\bibitem[{{Katsukawa} \& {Orozco Su{\'a}rez}(2012)}]{katsukawaorozco12}
{Katsukawa}, Y. \& {Orozco Su{\'a}rez}, D. 2012, \apj, 758, 139

\bibitem[{{Khomenko} {et~al.}(2005){Khomenko}, {Shelyag}, {Solanki}, \&
  {V{\"o}gler}}]{khomenkoetal05}
{Khomenko}, E.~V., {Shelyag}, S., {Solanki}, S.~K., \& {V{\"o}gler}, A. 2005,
  \aap, 442, 1059

\bibitem[{{Lin} \& {Rimmele}(1999)}]{linrimmele99}
{Lin}, H. \& {Rimmele}, T. 1999, \apj, 514, 448

\bibitem[{{Lites} {et~al.}(2008){Lites}, {Kubo}, {Socas-Navarro}, {Berger},
  {Frank}, {Shine}, {Tarbell}, {Title}, {Ichimoto}, {Katsukawa}, {Tsuneta},
  {Suematsu}, {Shimizu}, \& {Nagata}}]{litesetal08}
{Lites}, B.~W., {Kubo}, M., {Socas-Navarro}, H., {et~al.} 2008, \apj, 672, 1237

\bibitem[{{Matloch} {et~al.}(2010){Matloch}, {Cameron}, {Shelyag}, {Schmitt},
  \& {Sch{\"u}ssler}}]{matlochetal10}
{Matloch}, {\L}., {Cameron}, R., {Shelyag}, S., {Schmitt}, D., \&
  {Sch{\"u}ssler}, M. 2010, \aap, 519, A52

\bibitem[{{Matthews} {et~al.}(1995){Matthews}, {Proctor}, \&
  {Weiss}}]{matthewsetal95}
{Matthews}, P.~C., {Proctor}, M.~R.~E., \& {Weiss}, N.~O. 1995, Journal of
  Fluid Mechanics, 305, 281

\bibitem[{{Meneguzzi} \& {Pouquet}(1989)}]{meneguzzipouquet89}
{Meneguzzi}, M. \& {Pouquet}, A. 1989, Journal of Fluid Mechanics, 205, 297

\bibitem[{{Moll} {et~al.}(2011){Moll}, {Pietarila Graham}, {Pratt}, {Cameron},
  {M{\"u}ller}, \& {Sch{\"u}ssler}}]{molletal11}
{Moll}, R., {Pietarila Graham}, J., {Pratt}, J., {et~al.} 2011, \apj, 736, 36

\bibitem[{{Muller} {et~al.}(1992){Muller}, {Auffret}, {Roudier}, {Vigneau},
  {Simon}, {Frank}, {Shine}, \& {Title}}]{mulleretal92}
{Muller}, R., {Auffret}, H., {Roudier}, T., {et~al.} 1992, \nat, 356, 322

\bibitem[{{Nagata} {et~al.}(2008){Nagata}, {Tsuneta}, {Suematsu}, {Ichimoto},
  {Katsukawa}, {Shimizu}, {Yokoyama}, {Tarbell}, {Lites}, {Shine}, {Berger},
  {Title}, {Bellot Rubio}, \& {Orozco Su{\'a}rez}}]{nagataetal08}
{Nagata}, S., {Tsuneta}, S., {Suematsu}, Y., {et~al.} 2008, \apjl, 677, L145

\bibitem[{{November} {et~al.}(1981){November}, {Toomre}, {Gebbie}, \&
  {Simon}}]{novemberetal81}
{November}, L.~J., {Toomre}, J., {Gebbie}, K.~B., \& {Simon}, G.~W. 1981,
  \apjl, 245, L123

\bibitem[{{Orozco Su{\'a}rez} {et~al.}(2007){Orozco Su{\'a}rez}, {Bellot
  Rubio}, {del Toro Iniesta}, {Tsuneta}, {Lites}, {Ichimoto}, {Katsukawa},
  {Nagata}, {Shimizu}, {Shine}, {Suematsu}, {Tarbell}, \&
  {Title}}]{orozcoetal07}
{Orozco Su{\'a}rez}, D., {Bellot Rubio}, L.~R., {del Toro Iniesta}, J.~C.,
  {et~al.} 2007, \apjl, 670, L61

\bibitem[{{Orozco Su{\'a}rez} {et~al.}(2012){Orozco Su{\'a}rez}, {Katsukawa},
  \& {Bellot Rubio}}]{orozcoetal12}
{Orozco Su{\'a}rez}, D., {Katsukawa}, Y., \& {Bellot Rubio}, L.~R. 2012, \apjl,
  758, L38

\bibitem[{{Ossendrijver}(2003)}]{ossendrijver03}
{Ossendrijver}, M. 2003, \aapr, 11, 287

\bibitem[{{Pietarila Graham} {et~al.}(2009){Pietarila Graham}, {Danilovic}, \&
  {Sch{\"u}ssler}}]{pietarilaetal09}
{Pietarila Graham}, J., {Danilovic}, S., \& {Sch{\"u}ssler}, M. 2009, \apj,
  693, 1728

\bibitem[{{Rieutord} \& {Rincon}(2010)}]{rieutordrincon10}
{Rieutord}, M. \& {Rincon}, F. 2010, Living Reviews in Solar Physics, 7, 2

\bibitem[{{Rieutord} {et~al.}(2010){Rieutord}, {Roudier}, {Rincon}, {Malherbe},
  {Meunier}, {Berger}, \& {Frank}}]{rieutordetal10}
{Rieutord}, M., {Roudier}, T., {Rincon}, F., {et~al.} 2010, \aap, 512, A4

\bibitem[{{Rincon} {et~al.}(2005){Rincon}, {Ligni{\`e}res}, \&
  {Rieutord}}]{rinconetal05}
{Rincon}, F., {Ligni{\`e}res}, F., \& {Rieutord}, M. 2005, \aap, 430, L57

\bibitem[{{Rogachevskii} \& {Kleeorin}(1997)}]{rogachevskiikleeorin97}
{Rogachevskii}, I. \& {Kleeorin}, N. 1997, \pre, 56, 417

\bibitem[{{S{\'a}nchez Almeida} \& {Lites}(2000)}]{sanchezlites00}
{S{\'a}nchez Almeida}, J. \& {Lites}, B.~W. 2000, \apj, 532, 1215

\bibitem[{{Schekochihin} {et~al.}(2005){Schekochihin}, {Haugen}, {Brandenburg},
  {Cowley}, {Maron}, \& {McWilliams}}]{schekochihinetal05}
{Schekochihin}, A.~A., {Haugen}, N.~E.~L., {Brandenburg}, A., {et~al.} 2005,
  \apjl, 625, L115

\bibitem[{{Sch{\"u}ssler}(2013)}]{schussler13}
{Sch{\"u}ssler}, M. 2013, in IAU Symposium, Vol. 294, IAU Symposium, ed. A.~G.
  {Kosovichev}, E.~{de Gouveia Dal Pino}, \& Y.~{Yan}, 95--106

\bibitem[{{Sch{\"u}ssler} \& {V{\"o}gler}(2008)}]{schusslervogler08}
{Sch{\"u}ssler}, M. \& {V{\"o}gler}, A. 2008, \aap, 481, L5

\bibitem[{{Shine} {et~al.}(2000){Shine}, {Simon}, \& {Hurlburt}}]{shineetal00}
{Shine}, R.~A., {Simon}, G.~W., \& {Hurlburt}, N.~E. 2000, \solphys, 193, 313

\bibitem[{{Simon} \& {Leighton}(1964)}]{simonleighton64}
{Simon}, G.~W. \& {Leighton}, R.~B. 1964, \apj, 140, 1120

\bibitem[{{Spruit}(1979)}]{spruit79}
{Spruit}, H.~C. 1979, \solphys, 61, 363

\bibitem[{{Stein} \& {Nordlund}(2006)}]{steinnordlund06}
{Stein}, R.~F. \& {Nordlund}, {\AA}. 2006, \apj, 642, 1246

\bibitem[{{Stenflo}(1973)}]{stenflo73}
{Stenflo}, J.~O. 1973, Solar Physics, 32, 41

\bibitem[{{Tobias} \& {Cattaneo}(2008)}]{tobias2008}
{Tobias}, S.~M. \& {Cattaneo}, F. 2008, Journal of Fluid Mechanics, 601, 101

\bibitem[{{Trujillo Bueno} {et~al.}(2004){Trujillo Bueno}, {Shchukina}, \&
  {Asensio Ramos}}]{trujilloetal04}
{Trujillo Bueno}, J., {Shchukina}, N., \& {Asensio Ramos}, A. 2004, \nat, 430,
  326

\bibitem[{{V{\"o}gler} \& {Sch{\"u}ssler}(2007)}]{voglerschussler07}
{V{\"o}gler}, A. \& {Sch{\"u}ssler}, M. 2007, \aap, 465, L43

\bibitem[{{Webb} \& {Roberts}(1978)}]{webbroberts78}
{Webb}, A.~R. \& {Roberts}, B. 1978, \solphys, 59, 249

\bibitem[{{Weiss}(1966)}]{weiss66}
{Weiss}, N.~O. 1966, Royal Society of London Proceedings Series A, 293, 310

\bibitem[{{Yelles Chaouche} {et~al.}(2011){Yelles Chaouche}, {Moreno-Insertis},
  {Mart{\'{\i}}nez Pillet}, {Wiegelmann}, {Bonet}, {Kn{\"o}lker}, {Bellot
  Rubio}, {del Toro Iniesta}, {Barthol}, {Gandorfer}, {Schmidt}, \&
  {Solanki}}]{yellesetal11}
{Yelles Chaouche}, L., {Moreno-Insertis}, F., {Mart{\'{\i}}nez Pillet}, V.,
  {et~al.} 2011, \apjl, 727, L30

\end{thebibliography}

\end{document}